\documentclass[12pt,preprint]{aastex}

\righthead{Spectrophotometry of Landolt Standard Stars}
\lefthead{Stritzinger {\em et al.}}

\begin{document}

\title{An Atlas of Spectrophotometric Landolt Standard Stars}

\author{Maximilian Stritzinger,$^{1,2}$
Nicholas B. Suntzeff,$^3$
Mario Hamuy,$^4$
Peter Challis,$^5$
Ricardo Demarco,$^6$
Lisa Germany,$^7$
A. M. Soderberg,$^8$}

\affil{$^1$Max-Planck-Institut f\"ur Astrophysik, Karl-Schwarzschild-Str.\ 1,
85741 Garching, Germany \\
$^2$Visiting Astronomer, Cerro Tololo Inter-American Observatory \\
$^3$Cerro Tololo Inter-American Observatory\altaffilmark{9}, Casilla 603, La Serena, Chile \\
$^4$Las Campanas Observatory, Carnegie Observatories, Casilla 601, La Serena, Chile \\
$^5$Harvard-Smithsonian Center for Astrophysics, 60 Garden Street, Cambridge, MA 02138, USA\\
$^6$Department of Physics and Astronomy, Johns Hopkins University, 3400
N. Charles St., Baltimore, MD 21218, USA \\
$^7$European Southern Observatory, Alonso de Cordova 3107, Vitacura Santiago, Chile \\
$^8$Caltech Institute of Technology, 1201 E. California Blvd, Pasadena, CA 91125, USA}

\email{stritzin@mpa-garching.mpg.de \\
nsuntzeff@noao.edu \\
mhamuy@lco.cl \\
pchallis@cfa.harvard.edu \\
demarco@pha.jhu.edu \\
lgermany@eso.org \\
ams@astro.caltech.edu}

\altaffiltext{9}{Cerro Tololo Inter-American Observatory, Kitt Peak
National Observatory, National Optical Astronomy Observatories,
operated by the Association of Universities for Research in Astronomy,
Inc., (AURA), under cooperative agreement with the National Science
Foundation.}

\begin{abstract}
We present CCD observations of 102 Landolt standard stars
obtained with the R-C spectrograph on the CTIO 1.5 m telescope.
Using stellar atmosphere models we have extended the flux
points to our six spectrophotometric secondary standards, in
both the blue and the red, allowing us to produce flux-calibrated
spectra that span a wavelength range from 3050~\AA\ to 1.1 \micron.
Mean differences between $UBVRI$ spectrophotometry computed using 
Bessell's standard passbands and Landolt's published photometry is 
found to be 1$\%$ or less. 
Observers in both hemispheres will find these 
spectra useful for flux-calibrating spectra and through the use of 
accurately constructed instrumental passbands be able to compute 
accurate corrections to bring instrumental magnitudes to any desired
standard photometric system (S-corrections).
In addition, by combining empirical and modeled spectra of the Sun, 
Sirius and Vega, we calculate and compare synthetic photometry 
to observed photometry taken from the literature for these three stars. 
\end{abstract}
\section{Introduction}

From dedicated follow-up observations of supernovae
(hereafter SNe) it has become clear that systematic
magnitude differences can exist between data sets obtained at different 
telescopes for the same event. These differences can be on the order 
of several hundredths of a magnitude or more near maximum light and  
potentiality larger for late-time photometry when the spectrum enters the 
nebular phase. This effect is undoubtedly caused by the use of filter sets 
employed at different telescopes, which do not exactly match each other 
\citep{suntzeff88,menzies89,hamuy90,suntzeff00}, and are magnified when 
the instrumental filters differ grossly from the standard
Johnson/Kron-Cousins passbands. Although the observed photometry is 
standardized to a common system through the use of color terms,
this is not expected to work perfectly, because there are
radical differences 
between the normal and continuous spectral energy distributions (SEDs) of the 
photometric standard stars compared to the SEDs of SNe, which are 
dominated by strong absorption and emission features.

Using SNe photometry uncorrected for this effect can lead to 
incorrect calculations of colors, host galaxy reddening, absolute magnitudes, 
and can 
ultimately bias cosmological parameters. However, the photometrist
may remedy this by computing ``S-corrections" to correct their 
photometry to a standard filter transmissivity function. 
An at least partially successful
attempt to reconcile these magnitude differences in the optical for the well 
observed SN~1999ee was made by \citet{stritzinger02}. More recently, this 
photometric technique has been used in the optical and extended to 
near infrared photometry by \citet{krisciunas03},~ \citet{candia03}, 
\citet{krisciunas04}, and \citet{pignata04} for a number of other well 
observed SNe. 

Spectrophotometric standard stars play a crucial role in determining accurate 
S-corrections. 
However, there exists only a small number of moderately faint 
standard stars --which are of limited color range-- 
useful for spectroscopic calibrations. 
In this work we construct a large atlas of flux-calibrated 
spectra in order to enlarge the hitherto available spectrophotometric 
standards. Our program consists of a large number ($\sim$100) of Landolt 
standard stars, which have well-documented photometric magnitudes and
are widely used for photometric calibrations. 
These standard stars now may be employed to flux-calibrate spectra necessary 
for determining many physical parameters of stars, (e.g. surface 
temperatures, radial velocities, abundances, surface gravities, etc.)
relate spectral and photometric observations, and to calculate $UBVRIz$-band 
S-corrections for any celestial object whose SED significantly differs from 
the standards used to calibrate the observed photometry.
These spectra are now available for other researchers, provided in electronic 
form as FITS files.\footnote[10]{At http://csp1.lco.cl/~mhamuy/SPECSTDS/}

The motivation of the authors for this study was to be able to model a
typical night's run of photometry at a facility telescope and CCD
instrument. We would like to be able to start with the SEDs of the
program objects, usually SNe, and the Landolt stars. Then, with
system transmission functions, which include atmospheric extinction,
mirror reflectivities, filter functions, dewar windows, and the
detector quantum efficiencies, calculate synthetic magnitudes as close
as possible to the observed natural system. 
Finally, we want to use the synthetic natural system magnitudes and
run them through our photometric codes to calculate the typical
extinction and color terms that are solved for each night. By comparing
the synthetic to the observed transformations, we can assess the
effects of many possible systematic errors in our data. How close do
the color terms match? How does the changing extinction across the
photometric bands affect the calculated colors at higher airmass? 
It should also be possible, in principle, to calculate the transmission
functions of the photometric bands from scratch using the observed
photometric solutions and the SEDs \citep{jha02}. 
\citeauthor{jha02} modeled the transmission curves with cubic splines
spaced equally over the wavelength region where a non-zero response
was expected in the $UBVRI$-bands. Typically six to eight spline points were
used in each band with the first and last points forced to zero at
wavelengths expected to have zero transmissivity slightly outside the
expected passbands. Roughly twenty spectrophotometric standards with
well established $UBVRI$ magnitudes were used in creating the synthetic
magnitudes, and were observed with the filter system in his study. A
best fit model transmission function in each filter was made by
minimizing the residuals between the synthetic and observed
magnitudes. The model fits the amplitudes of the spline points
restricted to values between zero and one. \citeauthor{jha02} noted
that the model
functions reproduce the transmission functions measured in the lab
reasonably well. Increasing the number of standards observed would
improve the fits, and possibly allow for more spline knots to be used.

As an example, which may surprise most astronomers who do not do
photometry, very few photometrists use second-order (color dependent)
terms in the extinction because it is difficult to measure this effect
accurately. Most extinction is handled as a simple grey shift of the
form $m(nat)_0 = m(nat) - k*X$, where $X$ is the extinction and $k$ is
the extinction coefficient of a given bandpass.  
The extinction curve is included in
our system throughput curves when doing synthetic photometry, but we
do not gauge its effects on the color for stars observed at $X=1$
versus $X=2$. With this atlas of Landolt SED spectra it is now 
possible to calculate the second order terms using synthetic photometry.

The structure of this paper is as follows. 
In $\S$~\ref{obs.sec}
we present our observations, followed by the spectroscopic reductions
in $\S$~\ref{sr.sec}. Our results for the program stars are given in 
$\S$~\ref{results}. 
Finally, in $\S$~\ref{results}
synthetic photometry of the Sun, Sirius and Vega is computed and compared
to observed magnitudes found in the literature.
\section{Observations}
\label{obs.sec}

Six bright (4.3 $\la$ $V$ $\la$ 5.7) secondary standard
stars (see Table \ref{secstand}) originally published in \citet{hayes70}, and 
later re-calibrated by \citet{taylor84,hamuy92,hamuy94} were adopted
as our defining spectrophotometric system. These stars are 
secondary standards because they tie the Kitt Peak National 
Observatory and CTIO spectrophotometric standards 
\citep{massey88,hamuy92,hamuy94} to Vega.
The program consists of 102 Landolt standard stars located along 
the celestial equator ranging 7.0 $\la$ $V$ $\la$ 13.0.
The reader is referred to \citet{landolt83,landolt92a,landolt92b},
\citet{hamuy92}, and references within for spectral classifications,
$UBVRI$-band photometry, coordinates, and finding charts. The reference
and observed photometry for each star may also be
found in the image header of each spectrum.

All observations were obtained with the CTIO 1.5 m, using the R-C
spectrograph, during 5 - 13 February 1999 (UT). Of the eight nights 
observed, all 
were photometric except the last night of 12 - 13 February 1999 (UT).
Half of the observations were dedicated to a blue setup while the other half 
were allocated to a red setup. The blue setup employed a low dispersion 
grating (300 lines mm$^{-1}$) with a dispersion of 2.85~\AA\ per pixel 
blazed at 4000~\AA\ and a 1200 $\times$ 800 LORAL CCD.
We observed in first order with a total wavelength coverage of
3300~\AA\ (3100 - 6400~\AA) and a FWHM resolution of 8.6~\AA.
The red setup consisted of a low dispersion grating (158 lines mm$^{-1}$)
with a dispersion of 5.34~\AA\ per pixel blazed at 8000~\AA\ with the
same LORAL CCD. We observed in first order with a total wavelength
coverage of 4800~\AA\ (5800 - 10,600~\AA) and a FWHM resolution of 16.4~\AA.
A OG570 second order blocking filter was used to suppress any leakage, which
would have otherwise contaminated the spectra red-ward of 6000~\AA.

Daily observations began with obtaining calibration images. This included
bias frames, dome flats with a 2$\arcsec$ and
21$\arcsec$ slit and finally twilight flats through a 21$\arcsec$ slit.
Nights in which we observed with the blue setup, projector flats were taken
with a quartz lamp.
With the projector flats we used 
CuS0$_{4}$ and Corning 9863 filters to prevent saturation of 
the CCD. The observing procedure consisted of (1) pointing the
telescope to the coordinates of the standard star, (2) close the slit
to 2$\arcsec$ and then take an exposure with a HeAr lamp, (3) 
select a random field star for telescope guiding purposes, and (4)
take an exposure of the standard star with a slit width of 21$\arcsec$. 
On the first two nights the 
secondary standards were each observed at three slit positions and the 
program stars at two slit positions. By the third night {\it all} stars were 
observed with four slit positions.\footnote[11]{A preliminary data reduction 
showed that all of the spectra from 5800 to 7000~\AA\ were choppy 
at the 2-4$\%$ level. This choppiness was similar to broad-scale
fringing which is typically seen at wavelengths longward of 8000~\AA.
To alleviate this problem, the observing procedure was changed 
to observe all stars at four slit positions at lower flux levels in order to 
obtain similar total integration times. The co-added frames reduced the level
of choppiness by half.}

For each night typically five or six secondary standard stars
were observed
(see Table \ref{secstand}), obtaining between 50 and 70
spectra at a wide range of airmass between $X = 1.0$ and $X = 2.3$,
in order to solve for the nightly extinction curve.
When observing the program stars
we restricted the range of airmass to between $X = 1.0$ and $X = 1.3$
in order to reduce the differential effects of the Earth's 
atmosphere such as telluric absorption and atmospheric refraction
between the program and spectrophotometric standard stars.
Integration times were chosen such
that for the majority of 1-D spectra (resulting from adding all the flux
in the 2-D image along the spatial direction) the number of counts
was between 40,000 and 50,000 ADU per resolution element. 
For all observations the predicted gain of the LORAL CCD was 1.420 detected
electrons per ADU.
Exposure times for the bright secondary standards
ranged between 2 and 7 seconds, while exposure times for
the program stars typically ranged between 25 and 400 seconds. Due to the
short integration times of the secondary standards it proved necessary to
apply a shutter correction to their spectra (see below $\S$\ref{sr.sec}). 
From multiple exposures taken
with 1-s, 2-s , 3-s, 4-s and 6-s exposure times on 6 - 7 February 1999 (UT) an
additive mean shutter error for a one second exposure was determined to
be -0.023 seconds $\pm$0.010 (s.d.).
\section{Spectroscopic Reductions}
\label{sr.sec}

Standard spectroscopic reduction techniques using
IRAF\footnote[12]{The Image Reduction and Analysis Facility (IRAF) is
distributed by the National Optical Astronomy Observatory, which is
operated by AURA Inc., under a cooperative agreement with the National
Science Foundation.} were performed to reduce the data. To begin  
the over-scan and bias was subtracted from all spectra including the 
HeAr frames. With the blue setup a flat field image was constructed using
a combination of dome flats (external illumination), projector flats
(internal illumination), and sky flats. The projector flats provide
illumination in the ultraviolet end of the CCD ($\lambda$ $<$ 3800~\AA), the
dome flats at redder wavelengths, and the sky flat permitted us to 
correct the dome and projector flats for uneven illumination along the 
slit. With the red setup we only used dome and sky flats. The resulting
flats (normalized along the dispersion axis) were divided into all of the 
observed spectra. Next we extracted 1-D spectra
from the 2-D flat fielded images and dispersion-calibrated them to a
linear wavelength scale using the HeAr calibrations frames that were
taken before each exposure.
Shutter corrections were then applied to all the secondary standards
by multiplying a factor of
\begin{equation}
\frac{ET}{ET + ST}
\end{equation}
into each spectra, where ET is the requested exposure time in seconds and ST 
is the mean shutter error given in $\S$ \ref{obs.sec}.

If the program stars are to be used as spectrophotometric standards
for calculating $U$- and $z$-band spectrophotometry
it proved necessary to extend the wavelength range of their spectra
beyond the 3300 - 10,406~\AA\ range of the secondary standards.
This was accomplished by fitting synthetic spectra modeled with 
appropriate
physical parameters, to each of our secondary standards, using 
Robert Kurucz's stellar atmosphere code BILL.f.\footnote[13]{R. Kurucz's 
stellar atmosphere models can be downloaded from his website
http://kurucz.harvard.edu/}
Input parameters for the BILL.f program include surface temperature, log g, and
metallicity abundances. The output models of this program are in step-sizes of
10~\AA\ and have a FWHM resolution of $\sim$ 6~\AA. 
By extrapolating from the models we obtained six new flux points.
These included four flux points blue-ward of 3300~\AA\ at
3250, 3200, 3150 and 3100~\AA\, and two flux points red-ward of 10,406~\AA\ 
at 10,500 and 10,600~\AA. 
It was necessary to scale
the models to the observed blue and red spectra by multiplication of 
an arbitrary constant. This constant was derived such that the modeled 
spectrum could
reproduce the same values (up to two significant figures) as the
flux points given in Table~5 of \citet{hamuy92}.
In addition, because two of the \citet{hamuy94} flux points 
were placed in regions of strong atmospheric contamination,
they were removed. However to account for this wavelength interval
we placed two additional flux points at 7845 and 9915~\AA.
Also, two new flux points were added at 9970 and 10,150~\AA.
Because there are no flux points from 8376 - 9834~\AA\, we attempted
to add two flux points in this interval at locations free of
atmospheric and stellar absorption features,  
at 8800 and 8920~\AA, each with a 10~\AA\ bandwidth. 
Unfortunately, when deriving the nightly response curve these flux points 
showed systematic residuals up to $\sim$~0\fm10 compared to neighboring 
flux points and were thus omitted.

Table~\ref{secstand} lists the re-calibrated
monochromatic magnitudes of our spectrophotometric secondary standards
from 3100 to 10,600~\AA. These values are defined by
\begin{equation}
m_{\nu}=-2.5 {\rm log}_{10}[f_{\nu}]-ZP,
\end{equation}
where f$_{\nu}$ is the monochromatic flux in ergs
cm$^{-2}$ s$^{-1}$ Hz$^{-1}$, and $ZP$ is the zero-point for the
magnitude scale.
The zero-point of the monochromatic magnitude scale was chosen
to be -48.590 \citep{massey88}. In order to allow the reader to 
easily compare the monochromatic magnitudes listed in 
Table~\ref{secstand} with those provided in \citet{hamuy92} we 
list them in units of m$_{\nu}$. However, for the rest of the
paper we work in units of f$_{\lambda}$ rather than 
f$_{\nu}$.

Before flux calibrating the spectra it is important to remove as
much of the instrumental artifacts (such as fringing) and telluric
absorption as possible. The most difficult signature to remove 
is the flat-fielding error, which introduces very high-order
variations in the continuum at the few percent level, due to the
continuum fitting algorithms used to take a flat field lamp and
``flatten" it with an IRAF task like RESPONSE. Typically we used a
polynomial of order 20 to 30 to fit out the flat field response. This
will introduce wiggles with a period of roughly 150~\AA\ or so, which
are impossible to remove with a polynomial fit to the Hayes flux
points, which are often more than 200~\AA\ separated. In the region of
8000 - 9500~\AA\, the flux points are even more separated, and one cannot
fit out these flat-fielding errors.

However, \citet{bessell99} who noticed correlated errors in the Hamuy et
al. spectrophotometric standards, has suggested an ingenious way of
removing these flat-fielding errors. He proposed that the data be
divided by a spectral flat, preferably with a spectrum of an
astrophysical source that is close to a black body or otherwise line
free. There is no such source, but there are some stars listed in his
table such as Feige 110 or VMa2, which are close to being a pure
continuum source. Most of these stars were too faint for the 1.5 m so
we had to do the next best thing -- use the division of an observed
spectrum of a hot star with the model of the hot star. We used HR 3454, 
which was observed every night at an airmass $\sim$1.2.
To construct the red spectral flat we first made a theoretical spectrum
using the Kurucz code at the same dispersion and
wavelength coverage as the observed spectra.
Because the Kurucz models are only available for a
large grid of physical parameters, it was necessary to interpolate
from the models to produce a spectrum that most accurately
matched HR~3454.\footnote[14]{This spectrum corresponds to a model
produced for physical parameters
($T_{eff}$, log $g$, [M/H]) $=$ (18650 K, 3.5, 0).}
All the red dispersion-calibrated spectra of HR~3454 at 
an airmass of 1.2 were then averaged and divided by the modeled spectrum. 
A few of the strongest spectral
features, such as H$\alpha$, did not cleanly disappear in the spectral
flat. Therefore these residuals were removed by interpolation.
Next, all of the red dispersion-calibrated
data for each night were divided by this spectral flat field.
Fig.~\ref{specphot.fig1} displays the main telluric features red-ward of 
6000~\AA\ that were removed from all the red spectra by division of 
the spectral flat. The most prominent 
telluric features were those associated with atmospheric H$_{2}$O and O$_{2}$.
The O$_{2}$ A- and B-bands were saturated for all observations, 
whereas the strength of the H$_{2}$O features were strongly dependent on 
both the airmass and the time at which the star was observed.

The blue spectral flat was constructed by 
averaging all the dispersion-calibrated observations of HR~3454 made at an 
airmass of 1.2. The blue spectral flat was then divided into all of the 
blue dispersion-calibrated data. Through the use of spectral flat fields 
we obtained smooth dispersion-calibrated spectra free of large telluric 
absorption and instrumental features.

Next we proceeded to flux-calibrate the data using the our fully 
reduced spectrophotometric secondary standards. Data from each night was first
corrected for atmospheric extinction, via the nightly extinction curves
derived from the secondary spectrophotometric standards, which were observed
to this end over a range of airmasses. 
In Fig.~\ref{specphot.fig2} we present an averaged extinction curve
obtained from the seven photometric nights for both the blue and the
red setups, and Table~2 lists this extinction curve in tabular 
form.
To obtain flux-calibrated spectra, a nightly response curve was 
derived by fitting a low order cubic spline to the observed flux values 
obtained from the secondary standards. When deriving nightly response curves 
we were able to extend the wavelength range past our reddest
flux point given in Table \ref{secstand} by 400~\AA\ to 11,000~\AA\,
and by 50~\AA\ in the blue to 3050~\AA.
To calculate spectrophotometry for each star it was necessary
to stitch the blue and red spectra together. This was accomplished by 
first comparing all the blue and red spectra for an individual star 
at an airmass of 1.5 or less. 
All of the spectra were combined 
by averaging them together (using the IRAF task scombine with 
the rejection option set to employ the averaged sigma clipping algorithm)
to produce a master spectrum for each 
star. Each master spectrum covers a total wavelength range of 
7950~\AA\ (3050 - 11,000~\AA). When considering all of individual 
spectra together the flux offsets were typically extremely small, 
on the order of $\sim$ 0.001 mag.

An indication that the Bessell method of using spectral flats works
well is that the flux calibrations curves were fit with lower ordered
polynomials. Without the division of the spectral flats, there
would have been 
noticeable wiggles in the sensitivity curve that would have required 
the use of high order
(12 or so) polynomials. With the division of the spectral flats, we
could use much lower polynomial fits. This gives us confidence that
using the spectral flats and fitting lower order polynomials, 
we have removed the systematic errors as seen by \citet{bessell99} in
the \cite{hamuy94} data.
\section{Results}
\label{results}

\subsection{Program Stars} 
In this section we want to assess the spectrophotometric
properties of our spectra by comparing broad-band synthetic
magnitudes to those measured by Landolt.
As all objects were measured with a photon detector, a synthetic
magnitude on the natural system must be calculated as the convolution
of a star's photon flux ($N_\lambda$) with the
filter instrumental passband (S($\lambda$)), i.e.
\begin{equation}
{\rm mag} = -2.5~{\rm log}_{10}~\int N_{\lambda}~S(\lambda)~d\lambda ~+~ZP,
\label{mageqn}
\end{equation}

\noindent where ZP is the zero-point for the magnitude scale.
The variable, S($\lambda$), should include the transparency of the
Earth's atmosphere, the filter transmission, the quantum efficiency (QE) 
of the detector, and mirror reflectivities. 

There is often confusion about the form of S($\lambda$). Some
references use a function of the form R($\lambda$)=$\lambda$*S($\lambda$) 
and integrate R($\lambda$)*F($\lambda$), where F($\lambda$) is in units of
ergs s$^{-1}$ cm$^{-2}$ \AA$^{-1}$. 
In Eq.~\ref{mageqn} we are specifying the photon flux in
units of photons s$^{-1}$ cm$^{-2}$ \AA$^{-1}$. With this definition, 
the meaning of S($\lambda$) is very easy to understand -- 
it is just the fraction of
photons (or energy) that is detected with respect to the incident flux
outside the earth's atmosphere. S($\lambda$) accounts for all the flux
lost due to the flux passing though the atmosphere, telescope, and
instrument.

To construct the standard passbands we adopted the 
Johnson/Kron-Cousins $UBVRI$ transmission functions given in \citet{bessell90}
(see our Fig.~\ref{specphot.fig3}).
Note, however, that the Bessell transmission functions are intended for use 
with energy 
rather then photon distributions (see Appendix in \citealt{bessell83}). Thus, 
it was necessary to divide these functions by $\lambda$ before employing them 
in Eq.~(\ref{mageqn}) \citep{suntzeff99,hamuy01}. 
Because telluric absorption features were removed from the spectra, an
atmospheric opacity spectrum was included in the construction of 
the standard passbands. 

Armed with the standard passbands we proceeded to calculate 
synthetic magnitudes for our spectra using zero-points determined 
from secondary spectrophotometric standards 
\citep{landolt92b,hamuy94,landolt99} rather than 
Vega, which has uncertain $UBVRI$-band optical photometry.  
When calculating zero-points we did not include telluric absorption
in the passbands because \citeauthor{hamuy94} did not removed 
these features from their spectra. Table \ref{zpts}
lists the resulting zero-points. Table \ref{mag} lists the 
synthetic magnitudes for all standard
stars computed with Eq.~(3) and the standard passbands, as 
well as the difference between observed and synthetic magnitudes. 
Note that the optical photometry for the secondary standards was taken from
\citet{hamuy92}.
Sufficient wavelength coverage was obtained for 98 of the 108 standard 
stars listed in Table \ref{mag} to calculate $UBVRI$ magnitudes. The reminding 
ten stars were observed
in either the blue or red except HD57884 and HD60826 whose spectra
were cut off blue-ward of 4000~\AA. In Table \ref{mag} we also identify stars
that are known or thought to be variable stars. 

In Fig.~\ref{specphot.fig4} we present, for all standards observed, the 
difference between observed  and synthetic magnitudes computed with the 
standard passbands, as a function of observed color. Overall there is a high 
internal accuracy between the observed and synthetic magnitudes as seen in 
Table \ref{mean}, which lists the mean difference and associated standard 
deviation for each band. Mean difference between the observed
photometry and our $UBVRI$-band synthetic magnitudes are 1\% or less.
However, it is evident from Fig.~\ref{specphot.fig4} 
that slight color terms do exist, most notably in the $U$- $B$- and $R$-bands. 
This color dependence reflects a small mismatch between 
the Bessell functions and the standards Johnson/Kron-Cousins
system and/or a possible error in the fundamental spectrophotometric
calibration. To remedy this problem our approach consisted of 
applying wavelength shifts to the Bessell functions until we obtained 
a zero color dependence. Table~\ref{shifts} lists the resulting shifts. 
Although small compared to the Bessell bandwidths ($\sim$ 1000~\AA),
they have a non-negligible effect on the synthetic magnitudes, and the
shifted standard passbands can be considered the best models 
for the Johnson/Kron-Cousins system. In Table~\ref{ubvbands} we provide
our new modeled $UBV$ standard passbands. In Table \ref{riband}
the $R$ and $I$ standard passbands are listed.  
Note, that the $R$ and $I$ standard passbands include an atmospheric line 
opacity spectrum.
In Fig.~\ref{specphot.fig5} we present the comparison of 
Bessell's standard passbands (dotted lines) to our new modeled passbands
(dashed lines). In addition, to 
complement the shifted passbands we provide a convenient list of wavelength 
shifts (see Table~\ref{nicklist}) one would apply to the standard passbands 
in order to increase the color term by 0.01 mag mag$^{-1}$.
In Table~\ref{nicklist} the color term (for example the $B$-band)
are in the form of $B = zpt + b_{nat} + K (B-V)$. Also listed
is the color used in each color term. 

With the $V$-band spectrophotometry and dispersion-calibrated spectra we 
investigated the LORAL CCD's response for all the nights on which
observations were conducted. In Fig.~\ref{specphot.fig6} we present the 
difference between standard and synthetic $V$-band magnitudes
as a function of counts in the extracted 1-D spectra at the effective
$V$-band wavelength for all observations.
We conclude from Fig.~\ref{specphot.fig6} that the response function 
of the LORAL CCD was linear to within 2$\%.$ 

\subsection{The Sun, Sirius and Vega}

In addition to the selected Landolt standard stars in this work,
we have calculated spectrophotometry for the Sun, Sirius and Vega.
As there are no spectrophotometric standards in the infrared comparable 
to that in the optical, these objects can be useful to 
accurately characterize the modeled passbands when computing 
$JHK$-band S-corrections (see 
\citet{krisciunas04}). Spectra for these objects have been constructed using 
a combination of empirical and modeled data. The reader is referred to
Appendix A in \citet{krisciunas03} for a more detailed description of the
construction of these spectra; below we provide a brief summary for each of 
these stars.

Our solar spectrum combines empirical data from \citet{livingston91}
scaled to a solar model from the Kurucz Web site \citep{kurucz84} with
physical parameters (T$_{eff}$, log g, $v_{micro}$, mixing length/scale
height) $=$ 5777 K, 4.438, 1.5 km s$^{-1}$, 1.25.
For Vega we have adopted observational data from
\citet{hayes85}. His
data were combined with the Kurucz spectrum vega090250000p.asc5
with physical parameters (T$_{eff}$, log g, $v_{micro}$, mixing length/scale
height) $=$ 9550 K, 3.950, -0.5, 2 km s$^{-1}$, 0.
The Kurucz model was then scaled to match the flux points of 
\citet{hayes85}. The Sirius spectrum was constructed using the 
Kurucz model sir.ascsq5 scaled to force the synthetic V magnitude to equal 
the observed value of -1.430 \citep{bessell98}. 
Each of these spectra were convolved to 2~\AA\ and re-sampled to
1~\AA~per pixel.

To compute $UBVRI$ synthetic photometry we employed our new
modeled passbands (shown in Fig.~\ref{specphot.fig5}) 
and the zero-points listed in 
Table~\ref{zpts}. To calculate 
$JHK$-band synthetic magnitudes we constructed instrumental passbands,
(see Fig.~\ref{specphot.fig7}), 
which included information associated with the Las Campanas Observatory's 
1 m Henrietta Swope 
telescope where the \citet{persson98} infrared system was established. This 
includes \citeauthor{persson98} 
$J_{S}$, $H$, and $K_{S}$ filter transmissivities, a Rockwell NICMOS2
QE response function, two aluminum reflections, a Dewar window 
transmissivity, multiple reflections associated with optical elements 
within the C40IRC camera, and an atmospheric line opacity spectrum. 
Zero-points were calculated by forcing the synthetic magnitudes of Vega to 
equal that of the \citet{elias82} CIT photometric system, 
i.e. ($J,H,K$) $=$ (0, 0, 0). 
The resulting $JHK$-band zero-points were -11.954, -11.895, and -12.063,
respectively.

Table \ref{sun} lists the published photometry (from multiple 
sources), our synthetic photometry and the difference between 
the two in the sense of observed minus synthetic. The difference 
between Vega's $V$-band observed and synthetic magnitudes shows that our 
zero-points, calculated using the secondary standards from 
\citet{hamuy94}, have an associated error 
$\sim$ 0.01 mag. 
The large differences in the $UB$-bands may be due to the 
difficulty in obtaining accurate measurements of a star as 
bright as Vega.
There is poor agreement for the Sun between $UJHK$ observed and 
synthetic magnitudes. This as well is not surprising considering the 
difficulty in obtaining precise photometry of a source as bright 
and extended as the Sun. Some of the large $U$-band difference may be a 
result of the large variability of both the Sun's flux in the ultraviolet 
and Earth's atmospheric transmissivity. The differences in the 
infrared may be attributed to telluric absorption features that were not 
sufficiently accounted for in our manufactured instrumental passbands. 
The near infrared 
spectrophotometry of Sirius matches well with observed photometry 
to within 1\% or less, while in the optical the difference is 
on the order of 4\% or less.

\acknowledgments
We thank the anonymous referee for helpful comments.
A special thanks goes to Arlo Landolt for providing us updated
values for the spectrophotometric standards. We also
acknowledge Mike Bessell, Kevin Krisciunas, Brian Schmidt,
Eric Persson, and Fiorella Castelli for helpful discussions on
photometry and spectrophotometry. We thank Bruno Leibundgut 
for useful comments on a preliminary draft of this manuscript.
M.S. acknowledges financial support from HST grant GO-07505.02-96A, 
and the International Max-Planck Research School on Astrophysics. 
This research has made use of the SIMBAD database, 
operated at CDS, Strasbourg, France.

\clearpage

\figcaption[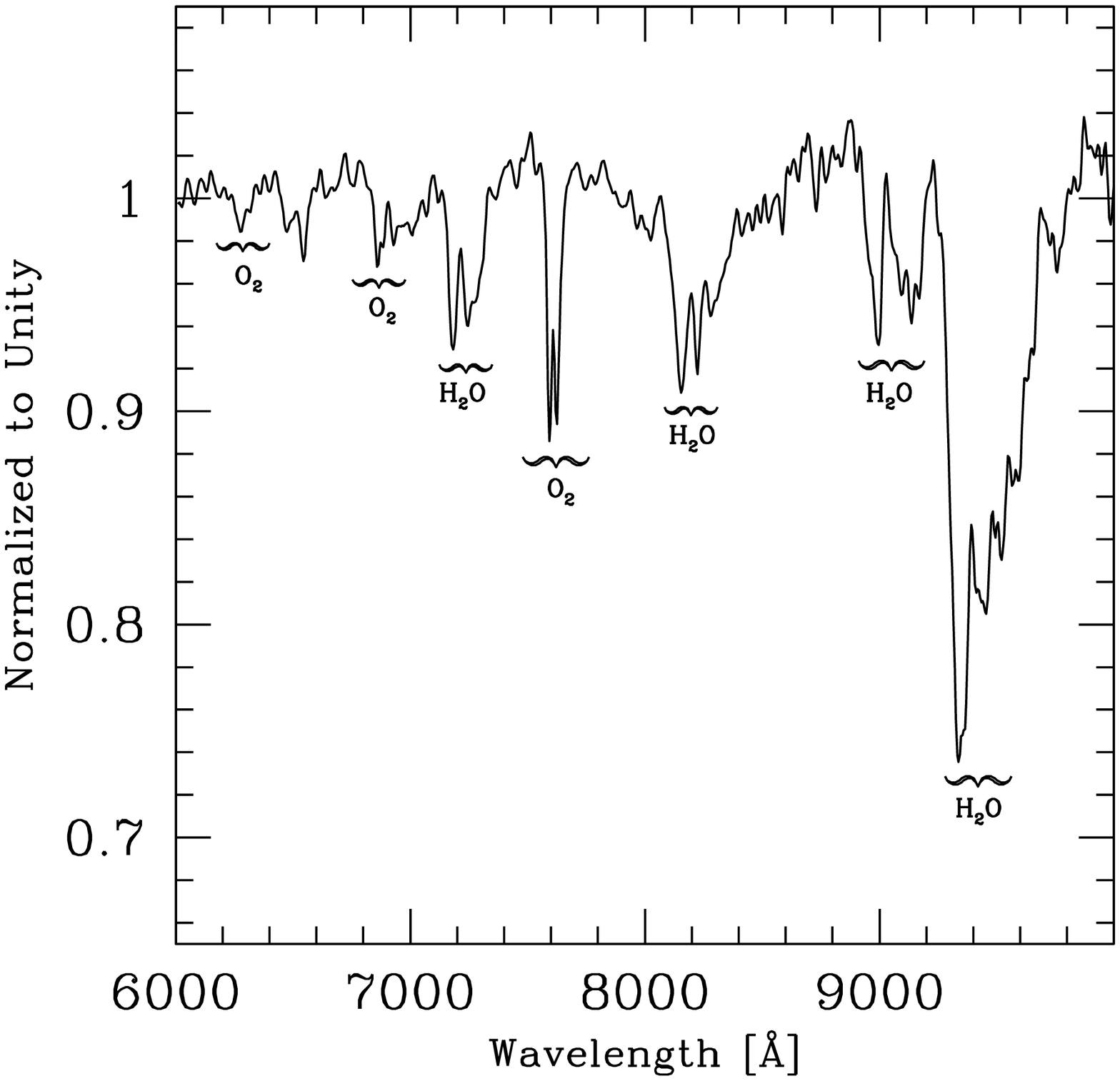]
{Plot of telluric features removed from all
spectra by division of a spectral flat. This figure is the division of
a high airmass spectrum by an intermediate airmass
spectrum of HR~3454 normalized to unity. The more prominent telluric features
are labeled. \label{specphot.fig1}}

\figcaption[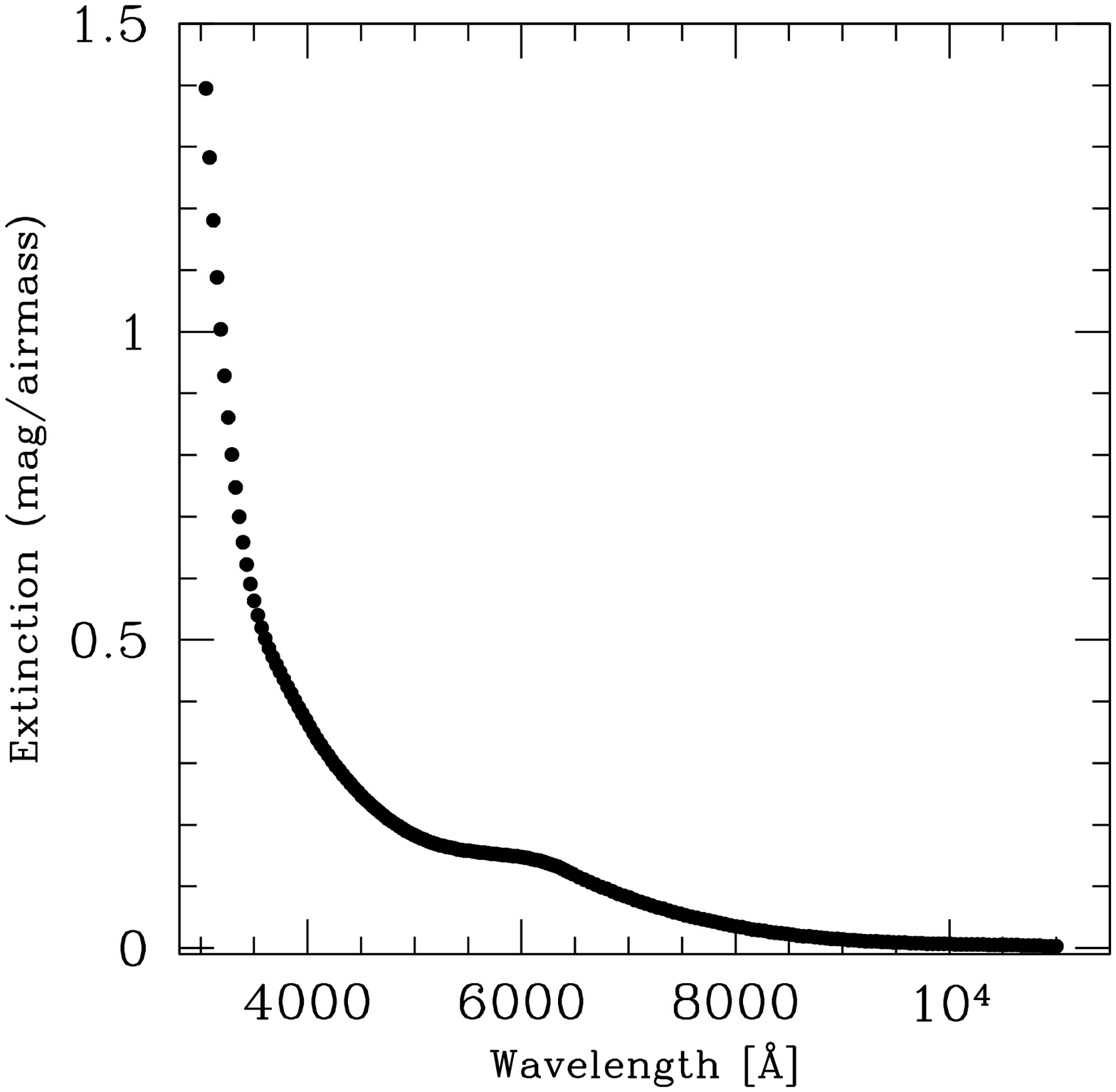]
{Averaged blue and red atmospheric extinction curves obtained at
CTIO on 5 - 12 February 1999 (UT). \label{specphot.fig2}}

\figcaption[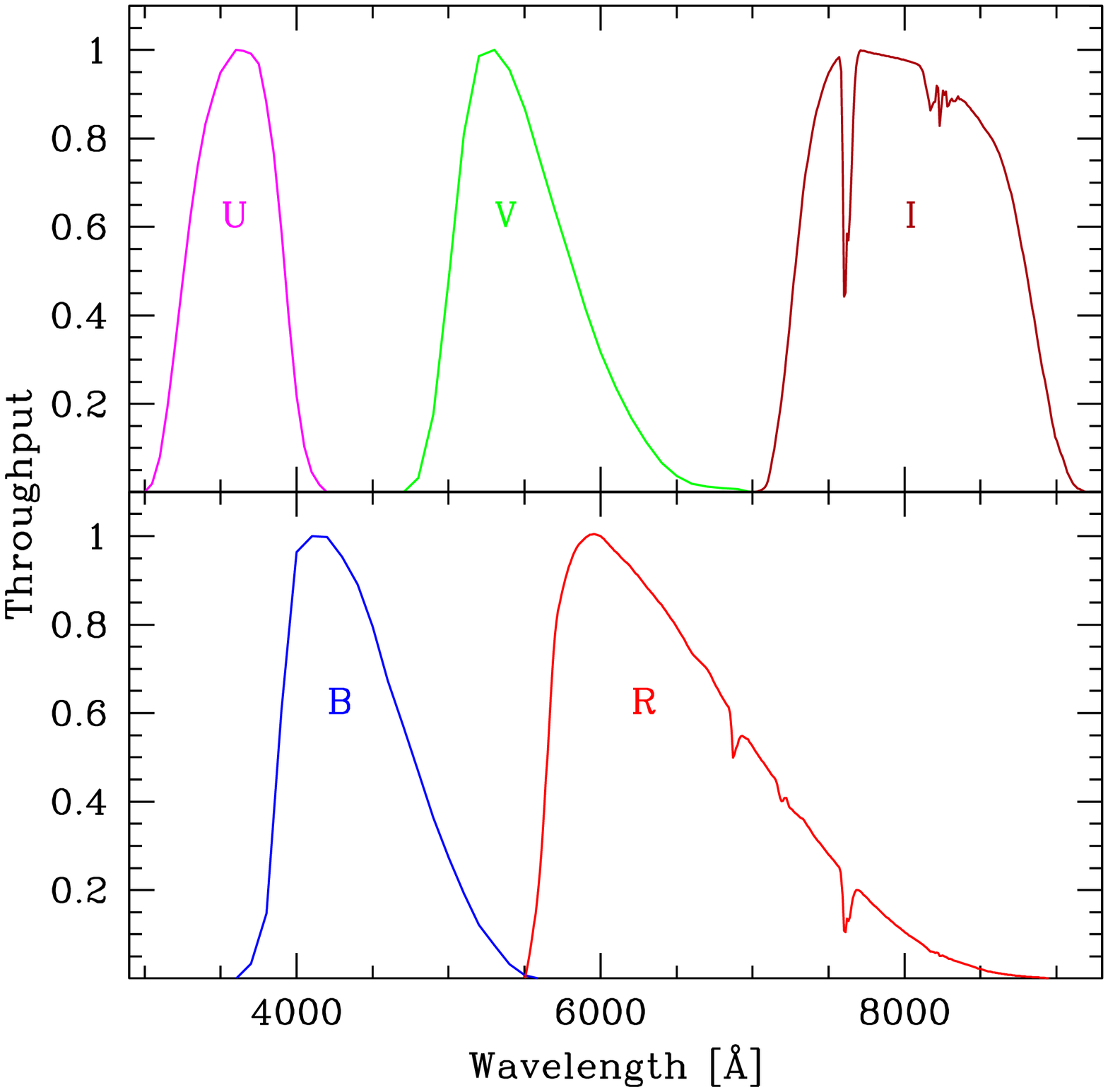]
{Johnson/Kron-Cousins $UBVRI$ standard passbands from
\citet{bessell90}. The Bessell transmission functions have been 
divided by $\lambda$ for integrations with photon flux, and multiplied 
by an atmospheric line opacity spectrum, because they are used 
with spectra that have had telluric features removed. \label{specphot.fig3}}

\figcaption[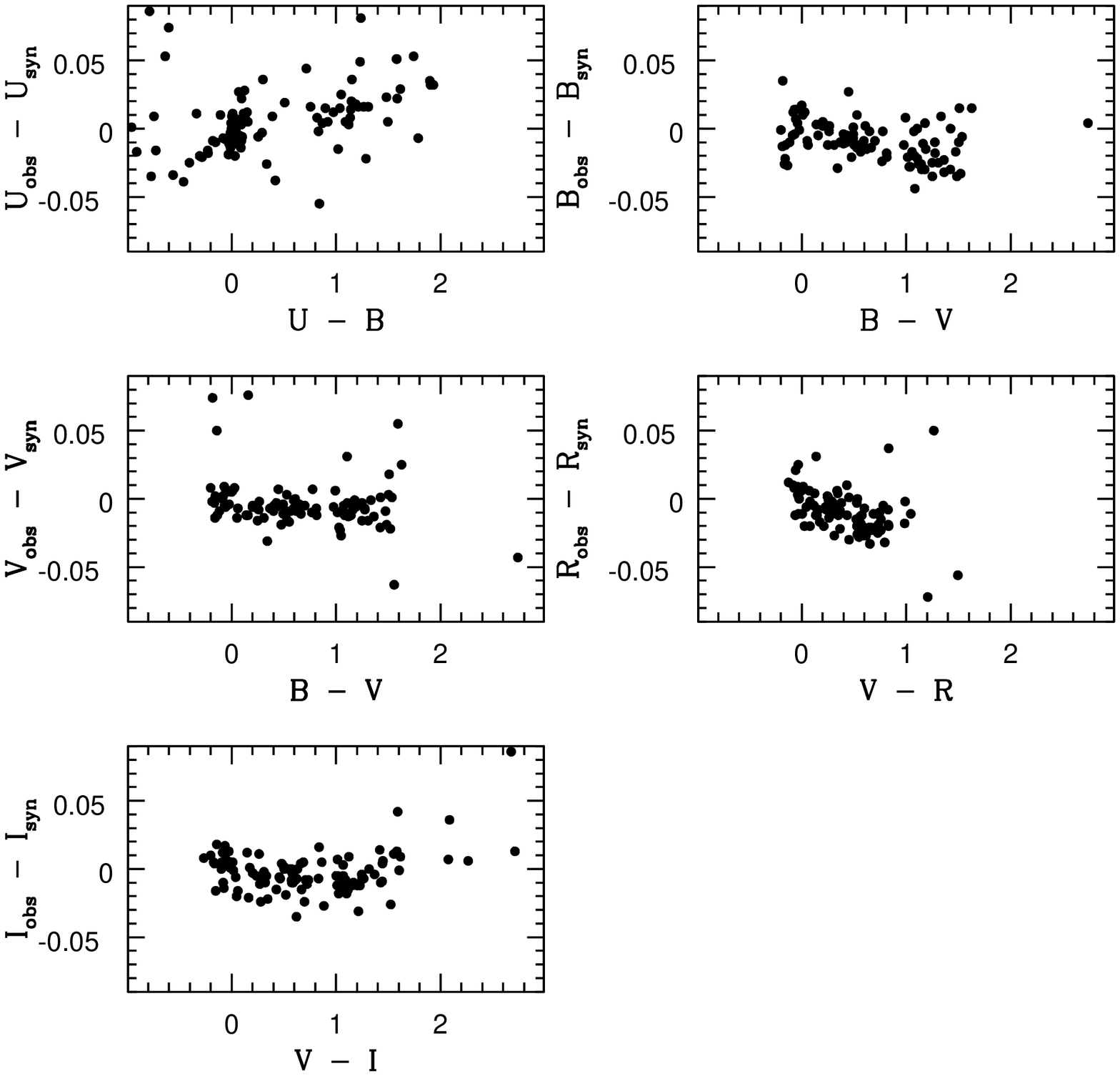]
{The difference between observed and synthetic magnitudes
derived using the Bessell
passbands (see Fig. \ref{specphot.fig3}) as a function of color. Significant
outliers are variable stars, identified in Table \ref{mag}. \label{specphot.fig4}}

\figcaption[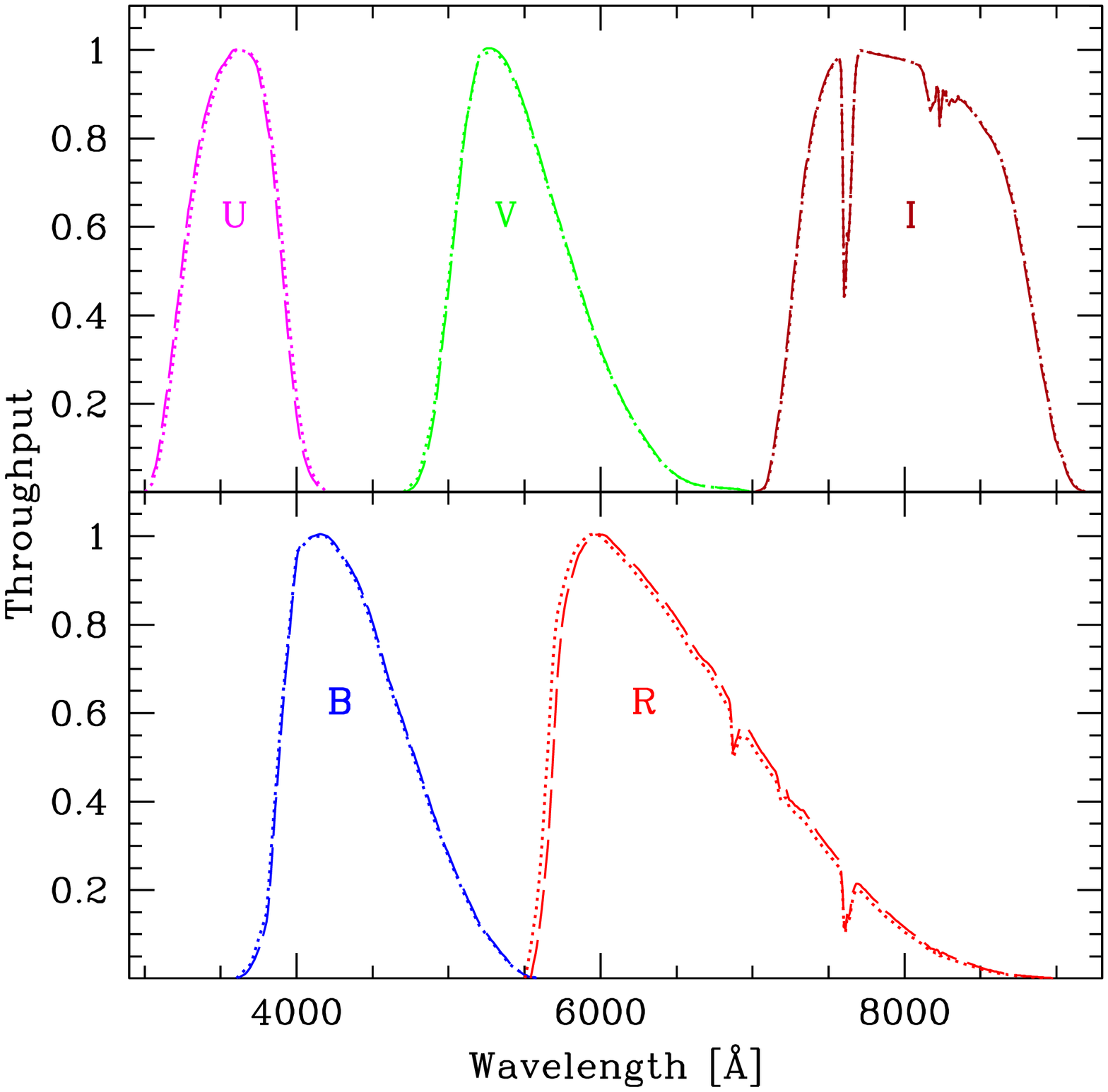]
{Comparison of the Bessell Johnson/Kron-Cousins passbands (dotted lines)
to our new modeled passbands (dashed lines) that include the shifts
listed in Table~\ref{shifts}. \label{specphot.fig5}}

\figcaption[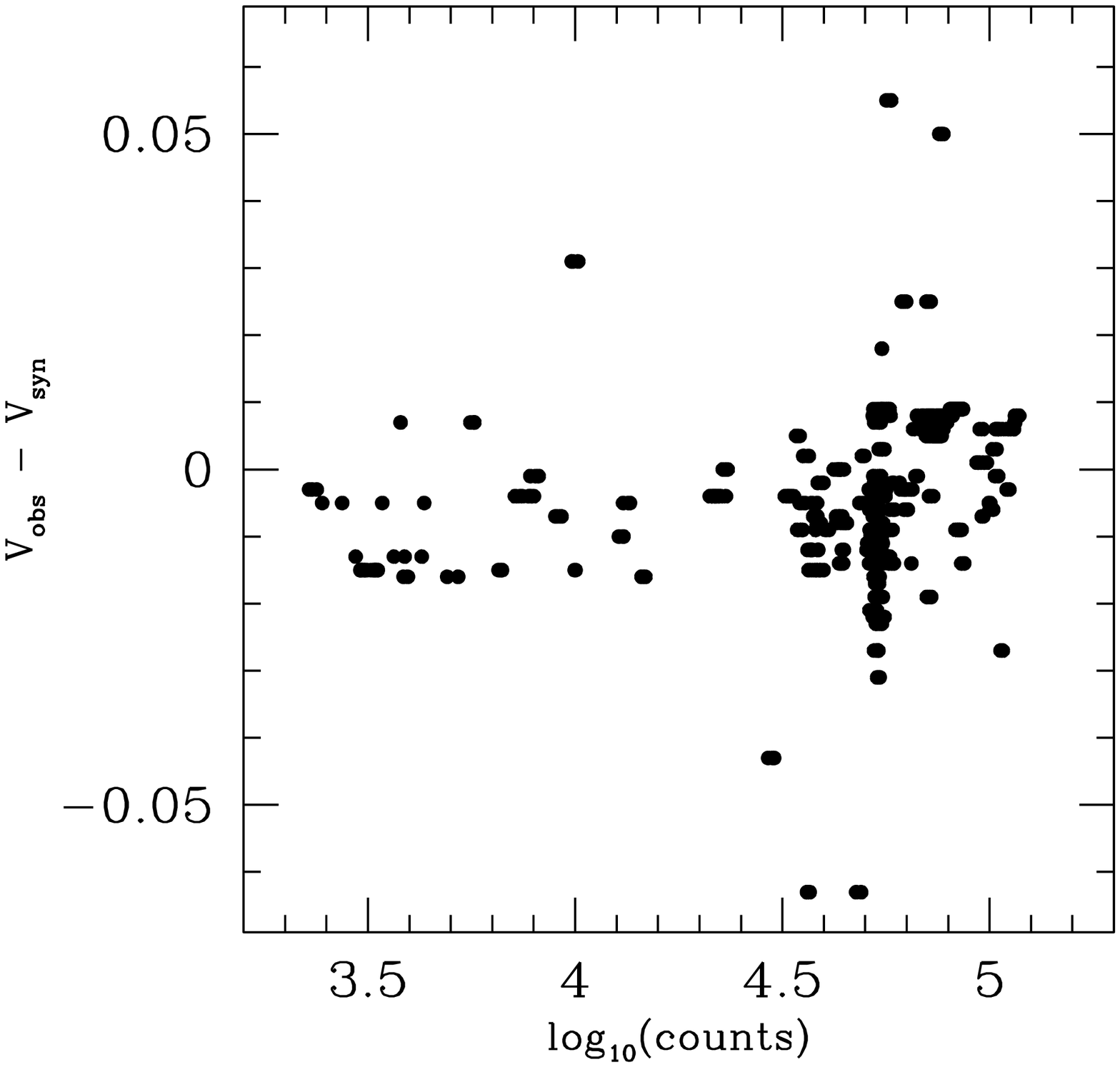]
{$V$-band observed minus synthetic magnitudes, verses
photon counts detected by the LORAL CCD, at the
$V$-band's effective wavelength. \label{specphot.fig6}}

\figcaption[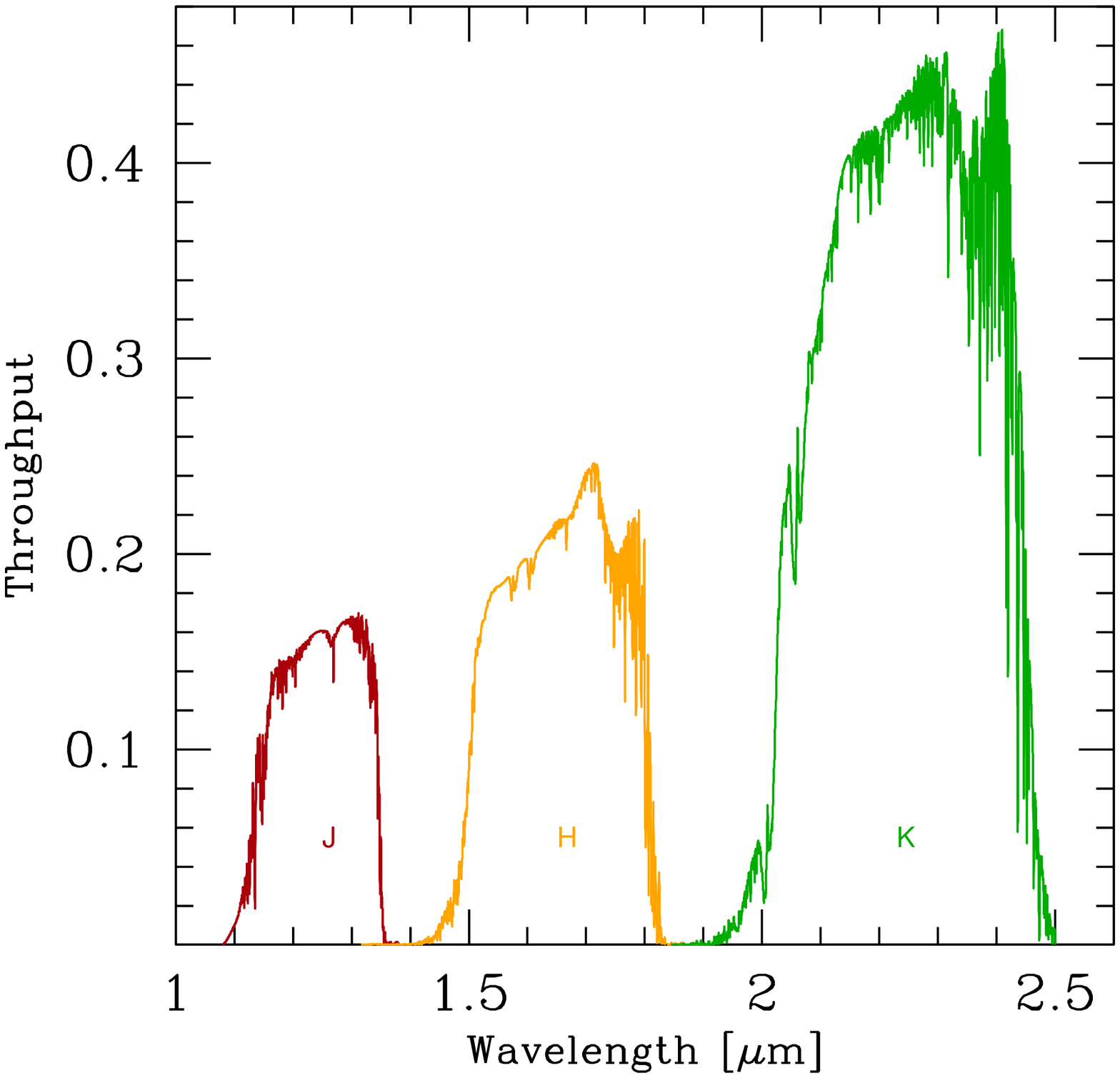]
{Infrared passbands corresponding to \citet{persson98}
$J_{S}$, $H$, and $K_{S}$ transmission functions, a NICMOS2
QE, multiple mirror reflections, a Dewar window transmissivity,
and an atmospheric line opacity spectrum. \label{specphot.fig7}}

\begin{figure}
\figurenum{1}
\epsscale{1.0}
\plotone{spdraft1.fig1.eps}
{\center Stritzinger {\it et al.} Fig. \ref{specphot.fig1}}
\end{figure}
\clearpage

\begin{figure}
\figurenum{2}
\epsscale{1.0}
\plotone{spdraft1.fig2.eps}
{\center Stritzinger {\it et al.} Fig. \ref{specphot.fig2}}
\end{figure}
\clearpage

\begin{figure}
\figurenum{3}
\epsscale{1.0}
\plotone{spdraft1.fig3.eps}
{\center Stritzinger {\it et al.} Fig. \ref{specphot.fig3}}
\end{figure}
\clearpage

\begin{figure}
\figurenum{4}
\epsscale{1.0}
\plotone{spdraft1.fig4.eps}
{\center Stritzinger {\it et al.} Fig. \ref{specphot.fig4}}
\end{figure}
\clearpage

\begin{figure}
\figurenum{5}
\epsscale{1.0}
\plotone{spdraft1.fig5.eps}
{\center Stritzinger {\it et al.} Fig. \ref{specphot.fig5}}
\end{figure}
\clearpage

\begin{figure}
\figurenum{6}
\epsscale{1.0}
\plotone{spdraft1.fig6.eps}
{\center Stritzinger {\it et al.} Fig. \ref{specphot.fig6}}
\end{figure}
\clearpage

\begin{figure}
\figurenum{7}
\epsscale{1.0}
\plotone{spdraft1.fig7.eps}
{\center Stritzinger {\it et al.} Fig. \ref{specphot.fig7}}
\end{figure}

\begin{deluxetable} {lccccccc}
\tabletypesize{\footnotesize}
\tablecolumns{8}
\tablenum{1}
\tablewidth{0pc}
\tablecaption{Spectrophotometric Secondary Standards \label{secstand}}
\tablehead{ 
\colhead{$\lambda$ [\AA]} & 
\colhead{$\Delta$ $\lambda$} &
\colhead{HR 718} &
\colhead{HR 1544} &
\colhead{HR 3454} &
\colhead{HR 4468} &
\colhead{HR 4963} &
\colhead{HR 5501} } 
\startdata   

3100  & 45 & 5.265 & 5.634 & 4.080 & 5.581 & 5.669 & 6.737 \\
3150  & 45 & 5.256 & 5.611 & 4.099 & 5.576 & 5.653 & 6.731 \\
3200  & 45 & 5.243 & 5.589 & 4.109 & 5.567 & 5.637 & 6.726 \\
3250  & 45 & 5.230 & 5.566 & 4.125 & 5.557 & 5.618 & 6.717 \\
3300  & 25 & 5.218 & 5.542 & 4.135 & 5.552 & 5.601 & 6.712 \\
3390  & 45 & 5.188 & 5.519 & 4.145 & 5.530 & 5.563 & 6.675 \\
3448  & 45 & 5.185 & 5.498 & 4.168 & 5.519 & 5.544 & 6.667 \\
3509  & 45 & 5.175 & 5.485 & 4.185 & 5.517 & 5.519 & 6.654 \\
3571  & 45 & 5.155 & 5.466 & 4.203 & 5.502 & 5.499 & 6.639 \\
3636  & 45 & 5.117 & 5.422 & 4.197 & 5.474 & 5.451 & 6.608 \\
4036  & 45 & 3.930 & 4.065 & 3.822 & 4.337 & 4.084 & 5.373 \\
4167  & 45 & 3.983 & 4.110 & 3.892 & 4.383 & 4.123 & 5.410 \\
4255  & 45 & 4.006 & 4.123 & 3.916 & 4.409 & 4.144 & 5.427 \\
4464  & 45 & \nodata   & 4.160 & 3.983 & 4.461 & 4.181 & 5.476 \\
4566  & 45 & 4.091 & 4.194 & 4.034 & 4.502 & 4.224 & 5.510 \\
4785  & 45 & 4.134 & 4.222 & 4.104 & 4.545 & 4.247 & 5.551 \\
5000  & 45 & 4.182 & 4.274 & 4.175 & 4.592 & 4.290 & 5.587 \\
5264  & 45 & 4.235 & 4.322 & 4.239 & 4.653 & 4.339 & 5.638 \\
5556  & 45 & 4.291 & 4.363 & 4.318 & 4.713 & 4.376 & 5.689 \\
5840  & 45 & 4.336 & 4.403 & 4.388 & 4.770 & 4.422 & 5.738 \\
6058  & 45 & 4.393 & 4.452 & 4.460 & 4.822 & 4.474 & 5.791 \\
6440  & 45 & 4.465 & 4.516 & 4.544 & 4.902 & 4.543 & 5.846 \\
6792  & 45 & 4.532 & 4.562 & 4.623 & 4.961 & 4.590 & 5.889 \\
7102  & 45 & 4.593 & 4.616 & 4.709 & 5.019 & 4.646 & 5.952 \\
7554  & 45 & 4.678 & 4.693 & 4.797 & 5.104 & 4.718 & 6.031 \\
7845  & 45 & 4.740 & 4.745 & 4.861 & 5.168 & 4.779 & 6.065 \\
8092  & 45 & 4.766 & 4.763 & 4.912 & 5.194 & 4.796 & 6.099 \\
8376  & 45 & 4.829 & 4.825 & 4.986 & 5.253 & 4.847 & 6.147 \\
8800  & 10 & 4.824 & 4.850 & 5.039 & 5.287 & 4.843 & 6.177 \\
8920  & 10 & 4.827 & 4.854 & 5.058 & 5.298 & 4.847 & 6.185 \\
9915  & 45 & 4.862 & 4.867 & 5.193 & 5.354 & 4.888 & 6.225 \\
9970  & 45 & 4.867 & 4.871 & 5.202 & 5.358 & 4.893 & 6.231 \\
10150 & 45 & 4.891 & 4.882 & 5.234 & 5.378 & 4.919 & 6.251 \\
10256 & 45 & 4.944 & 4.898 & 5.261 & 5.378 & 4.926 & 6.240 \\
10406 & 45 & 4.968 & 4.914 & 5.296 & 5.414 & 4.960 & 6.271 \\
10500 & 45 & 4.986 & 4.944 & 5.319 & 5.443 & 5.004 & 6.333 \\
10600 & 45 & 5.022 & 4.965 & 5.351 & 5.469 & 5.041 & 6.366 \\

\enddata 
\tablecomments{\scriptsize{All values are in monochromatic magnitudes m$_{\nu}$ = -2.5 log$_{10}$(f$_{\nu}$) - 48.590.}}  
\end{deluxetable}

\clearpage
\begin{deluxetable} {cc}
\tabletypesize{\footnotesize}
\tablecolumns{2}
\tablenum{2}
\tablewidth{0pc}
\tablecaption{Averaged Extinction Curve for CTIO \label{ext}}
\tablehead{
\colhead{$\lambda$} &
\colhead{Extinction [mag/airmass]}} 

\startdata
3050.00  &    1.395\\
3084.65  &    1.283\\
3119.31  &    1.181\\
3153.96  &    1.088\\
3188.61  &    1.004\\
3223.27  &   0.929\\
3257.92  &   0.861   \\
3292.57  &   0.801   \\
3327.23  &   0.748   \\
3361.88  &   0.700   \\
3396.54  &   0.659   \\
3431.19  &   0.623   \\
3465.84  &   0.591   \\
3500.50  &   0.564   \\
3535.15  &   0.540   \\
3569.80  &   0.520   \\
3604.46  &   0.502   \\
3639.11  &   0.487   \\
3673.76  &   0.473   \\
3708.42  &   0.460   \\
3743.07  &   0.448   \\
3777.72  &   0.436   \\
3812.38  &   0.425   \\
3847.03  &   0.414   \\
3881.69  &   0.402   \\
3916.34  &   0.391   \\
3950.99  &   0.381   \\
3985.65  &   0.370   \\
4020.30  &   0.360   \\
4054.95  & 0.349\\
4089.61  & 0.339\\
4124.26  & 0.330\\
4158.91  & 0.321\\
4193.57  & 0.313\\
4228.22  & 0.304\\
4262.87  & 0.296\\
4297.53  & 0.289\\
4332.18  & 0.281\\
4366.83  & 0.274\\
4401.49  & 0.267\\
4436.14  & 0.260\\
4470.79  & 0.254\\
4505.45  & 0.247\\
4540.10  & 0.241\\
4574.76  & 0.236\\
4609.41  & 0.230\\
4644.06  & 0.225\\
4678.72  & 0.220\\
4713.37  & 0.215\\
4748.02  & 0.210\\
4782.68  & 0.206\\
4817.33  & 0.202\\
4851.98  & 0.198\\
4886.64  & 0.194\\
4921.29  & 0.190\\
4955.94  & 0.187\\
4990.60  & 0.184\\
5025.25  & 0.181\\
5059.91  & 0.178\\
5094.56  & 0.176\\
5129.21  & 0.173\\
5163.87  & 0.171\\
5198.52  & 0.169\\
5233.17  & 0.167\\
5267.83  & 0.166\\
5302.48  & 0.164\\
5337.13  & 0.163\\
5371.79  & 0.162\\
5406.44  & 0.160\\
5441.09  & 0.159\\
5475.75  & 0.158\\
5510.40  & 0.158\\
5545.05  & 0.157\\
5579.71  & 0.156\\
5614.36  & 0.155\\
5649.02  & 0.155\\
5683.67  & 0.154\\
5718.32  & 0.153\\
5752.98  & 0.153\\
5787.63  & 0.152\\
5822.28  & 0.151\\
5856.94  & 0.151\\
5891.59  & 0.150\\
5926.24  & 0.149\\
5960.90  & 0.149\\
5995.55  & 0.148\\
6030.20  & 0.147\\
6064.86  & 0.146\\
6099.51  & 0.144\\
6134.17  & 0.143\\
6168.82  & 0.142\\
6203.47  & 0.140\\
6238.13  & 0.138\\
6272.78  & 0.136\\
6307.43  & 0.134\\
6342.09  & 0.132\\
6376.74  & 0.129\\
6411.39  & 0.126\\
6446.05  & 0.123\\
6480.70  & 0.120\\
6482.85  & 0.120\\
6535.38  & 0.115\\
6587.91  & 0.111\\
6640.44  & 0.107\\
6692.96  & 0.103\\
6745.49  & 0.099\\
6798.02  & 0.096\\
6850.55  & 0.092\\
6903.07  & 0.088\\
6955.60  & 0.085\\
7008.13  & 0.082\\
7060.65  & 0.078\\
7113.18  & 0.075\\
7165.71  & 0.072\\
7218.24  & 0.069\\
7270.76  & 0.066\\
7323.29  & 0.064\\
7375.82  & 0.061\\
7428.35  & 0.058\\
7480.87  & 0.056\\
7533.40  & 0.053\\
7585.93  & 0.051\\
7638.45  & 0.049\\
7690.98  & 0.047\\
7743.51  & 0.045\\
7796.04  & 0.043\\
7848.56  & 0.041\\
7901.09  & 0.039\\
7953.62  & 0.037\\
8006.15  & 0.035\\
8058.67  & 0.034\\
8111.20  & 0.032\\
8163.73  & 0.030\\
8216.25  & 0.029\\
8268.78  & 0.028\\
8321.31  & 0.026\\
8373.84  & 0.025\\
8426.36  & 0.024\\
8478.89  & 0.023\\
8531.42  & 0.022\\
8583.95  & 0.020\\
8636.47  & 0.019\\
8689.00  & 0.019\\
8741.53  & 0.018\\
8794.05  & 0.017\\
8846.58  & 0.016\\
8899.11  & 0.015\\
8951.64  & 0.015\\
9004.16  & 0.014\\
9056.69  & 0.013\\
9109.22  & 0.013\\
9161.75  & 0.012\\
9214.27  & 0.011\\
9266.80  & 0.011\\
9319.33  & 0.011\\
9371.85  & 0.010\\
9424.38  & 0.010\\
9476.91  & 0.009\\
9529.44  & 0.009\\
9581.96  & 0.009\\
9634.49  & 0.008\\
9687.02  & 0.008\\
9739.55  & 0.008\\
9792.07  & 0.007\\
9844.60  & 0.007\\
9897.13  & 0.007\\
9949.65  & 0.007\\
10002.2  & 0.007\\
10054.7  & 0.006\\
10107.2  & 0.006\\
10159.8  & 0.006\\
10212.3  & 0.006\\
10264.8  & 0.006\\
10317.3  & 0.006\\
10369.9  & 0.005\\
10422.4  & 0.005\\
10474.9  & 0.005\\
10527.5  & 0.005\\
10580.0  & 0.005\\
10632.5  & 0.005\\
10685.0  & 0.004\\
10737.6  & 0.004\\
10790.1  & 0.004\\
10842.6  & 0.004\\
10895.1  & 0.003\\
10947.7  & 0.003\\
11000.2  & 0.003\\

\enddata
\end{deluxetable}

\clearpage
\begin{deluxetable} {lc}
\tablecolumns{2}
\tablenum{3}
\tablewidth{0pc}
\tablecaption{Zero-points employed in Eq.~\ref{mageqn} with standard
passbands \label{zpts}}
\tablehead{
\colhead{Filter} &
\colhead{Zero-point} }
\startdata

$U$ & $-14.244$ \\
$B$ & $-15.279$ \\
$V$ & $-14.850$ \\
$R$ & $-15.053$ \\
$I$ & $-14.556$ \\

\enddata
\end{deluxetable}

\clearpage
\begin{deluxetable} {lcccccccccc}
\tabletypesize{\footnotesize}
\rotate
\tablecolumns{11}
\tablenum{4}
\tablewidth{0pc}
\tablecaption{Synthetic Magnitudes for All Stars$^\dag$ \label{mag}}
\tablehead{ 
\colhead{Star} & 
\colhead{$U$${\rm_{syn}}$} &
\colhead{$U$${\rm_{obs}}$-$U$${\rm_{syn}}$} &
\colhead{$B$${\rm_{syn}}$} &
\colhead{$B$${\rm_{obs}}$-$B$${\rm_{syn}}$} &
\colhead{$V$${\rm_{syn}}$} &
\colhead{$V$${\rm_{obs}}$-$V$${\rm_{syn}}$} &
\colhead{$R$${\rm_{syn}}$} &
\colhead{$R$${\rm_{obs}}$-$R$${\rm_{syn}}$} &
\colhead{$I$${\rm_{syn}}$} &
\colhead{$I$${\rm_{obs}}$-$I$${\rm_{syn}}$}}
\startdata   

bd$-$0\degr454   & 11.849 &$+0.051$ & 10.321 &  0     & 8.894  & $+0.001$ & 8.152  & $-0.015$ & 7.446  & $+0.004$ \\
bd$+$1$\degr$2447  & 12.318 &$+0.081$ & 11.146 & $+0.015$ & 9.634  & $+0.018$ & 8.619  & $-0.011$ & 7.380  & $+0.006$ \\
bd$+$5$\degr$1668$^c$&12.597&$+0.016$ & 11.460 & $-0.060$ & 9.906  & $-0.063$& 8.709  & $-0.072$ & 7.116  & $+0.013$ \\
bd$+$5$\degr$2468  & 8.706  &$-0.034$& 9.242  & $-0.010$& 9.357  & $-0.009$& 9.383  & $+0.003$& 9.435  & $+0.005$ \\
bd$+$5$\degr$2529  & 12.001 &$+0.018$ & 10.866 & $-0.035$& 9.584  & $-0.003$& 8.805  & $-0.005$& 8.125  & $+0.006$ \\
cd$-$32$\degr$9927 &\nodata &\nodata &\nodata &\nodata &\nodata &\nodata &\nodata &\nodata & 10.112 & $-0.005$ \\
eg21        &\nodata &\nodata &\nodata &\nodata &\nodata &\nodata &\nodata &\nodata & 11.526 & $+0.004$ \\
g162-66     & 11.861 & $-0.010$& 12.873 & $-0.026$& 13.015 & $-0.003$& 13.126 & $+0.012$& 13.270 & $+0.008$ \\
hd118246$^a$&7.259 & $+0.053$& 7.895  & $+0.053$& 8.039  & $+0.050$& 8.098  & $+0.025$& 8.180  & $-0.010$ \\
hd12021     & 8.413  & $-0.025$& 8.780  & $+0.012$& 8.872  & $+0.002$& 8.907  & $+0.009$& 8.972  &  0     \\
hd11983     &\nodata &\nodata &\nodata &\nodata &\nodata &\nodata &\nodata &\nodata & 6.627  & $+0.011$ \\
hd121968    & 9.177  & $-0.017$& 10.081 & $-0.013$& 10.256 & $-0.002$& 10.319 & $+0.008$& 10.421 & $+0.005$ \\
hd129975    & 11.741 & $+0.035$& 9.887  & $-0.010$& 8.370  & $+0.003$& 7.563  & $-0.019$& 6.769  & $-0.001$ \\
hd16581     & 7.853  & $-0.020$& 8.131  & $+0.007$& 8.201  & $-0.006$& 8.220  &  0     & 8.254  & $+0.004$ \\
hd21197     & 10.143 & $+0.014$& 9.046  & $-0.030$& 7.869  & $-0.003$& 7.191  & $-0.011$& 6.624  & $-0.004$ \\
hd36395     & 10.616 & $+0.049$& 9.451  & $-0.017$&  7.969 & $-0.009$& 6.993  & $-0.018$& 5.877  & $+0.007$ \\
hd47761$^b$&8.207  & $+0.074$ &8.798  & $+0.085$& 8.648  & $+0.076$& 8.555  & $+0.031$& 8.469  & $-0.024$ \\
hd50167     & 11.087 & $+0.053$& 9.402  & $-0.006$& 7.860  & $+0.001$& 7.043  & $-0.008$& 6.265  & $+0.013$ \\
hd52533     & 6.655  & $+0.001$& 7.619  & $-0.005$& 7.702  &  0     & 7.706  & $+0.007$& 7.734  & $+0.006$ \\
hd57884$^b$&\nodata&\nodata &\nodata &\nodata & 9.028  & $+0.107$& 7.821  & $+0.050$& 6.726  & $+0.122$ \\
hd60826$^b$&\nodata&\nodata &\nodata &\nodata & 9.026  & $-0.043$& 7.544  & $-0.056$& 6.220  & $+0.086$ \\
hd65079$^c$ &6.778  & $+0.086$& 7.615  & $+0.035$& 7.758  & $+0.074$& 7.763  & $+0.124$& 7.782  & $+0.180$ \\
hd72055     & 7.555  & $-0.039$& 8.003  & $-0.027$& 8.125  & $-0.012$& 8.142  & $+0.002$& 8.187  & $+0.012$ \\
hd76082     & 10.612 & $+0.005$& 9.553  & $-0.026$& 8.422  & $-0.013$& 7.851  & $-0.027$& 7.326  & $-0.018$ \\
hd79097$^a$ &11.130 & $+0.032$& 9.214  & $+0.015$& 7.576  & $+0.025$& 6.613  & $-0.002$& 5.478  & $+0.036$ \\
hd84971     & 7.742  & $-0.035$& 8.499  & $-0.022$& 8.650  & $-0.014$& 8.711  & $-0.012$& 8.804  & $-0.016$ \\
hd97503     & 10.998 & $+0.003$& 9.910  & $-0.030$& 8.703  & $-0.001$& 7.993  & $-0.011$& 7.385  &  0     \\
hr0718      & 4.106  & $+0.010$& 4.209  & $+0.014$& 4.272  & $+0.007$& 4.294  & $+0.008$& 4.331  & $+0.011$ \\
hr1544      & 4.372  & $-0.008$& 4.355  & $+0.010$& 4.349  & $+0.006$& 4.332  & $+0.009$& 4.322  & $-0.006$ \\
hr3454      & 3.343  & $+0.009$& 4.096  & $-0.001$& 4.287  & $+0.008$& 4.368  & $+0.010$& 4.485  & $+0.010$ \\
hr4468      & 4.459  & $-0.009$& 4.616  & $+0.014$& 4.691  & $+0.009$& 4.718  & $+0.005$& 4.746  & $+0.017$ \\
hr4963      & 4.379  & $-0.014$& 4.358  & $+0.017$& 4.370  & $+0.005$& 4.364  & $+0.008$& 4.360  & $+0.005$ \\
hr5501      & 5.585  & $-0.007$& 5.659  & $-0.001$& 5.685  & $-0.004$& 5.688  & $-0.011$& 5.694  & $+0.013$ \\
ltt1788     & 13.349 & $-0.021$& 13.618 & $-0.007$&\nodata &\nodata &\nodata &\nodata &\nodata &\nodata \\
ltt2415     & 12.388 & $-0.016$& 12.604 & $-0.004$&\nodata &\nodata &\nodata &\nodata &\nodata &\nodata \\
ltt3218     & \nodata&\nodata & \nodata&\nodata &\nodata &\nodata &\nodata &\nodata & 11.646 & $-0.003$ \\
ltt4364     & \nodata &\nodata&\nodata &\nodata &\nodata &\nodata &\nodata &\nodata & 11.189 & $-0.002$ \\
sa94-305    & 11.876 & $+0.022$& 10.342 & $-0.030$& 8.910  & $-0.021$& 8.155  & $-0.023$& 7.455  & $-0.009$ \\
sa94-308    & 9.229  & $+0.004$& 9.241  & $-0.004$& 8.754  & $-0.011$& 8.459  & $-0.006$& 8.177  & $-0.010$ \\
sa94-342    & 10.706 & $+0.044$& 10.026 & $+0.008$& 9.035  & $+0.006$& 8.517  & $-0.003$& 8.019  & $+0.007$ \\
sa95-52     & 10.145 & $+0.027$& 10.093 & $+0.010$& 9.571  & $+0.003$& 9.273  & $-0.006$& 8.971  & $-0.008$ \\
sa95-96     & 10.234 & $-0.005$& 10.154 & $+0.003$& 10.022 & $-0.012$& 9.933  & $-0.002$& 9.835  & $+0.001$ \\
sa95-132    & 12.776 & $+0.036$& 12.485 & $+0.027$& 12.057 & $+0.007$& 11.805 &   0    & 11.519 &  0     \\
sa95-206    & 9.259  & $-0.005$& 9.251  & $-0.012$& 8.748  & $-0.011$& 8.449  & $-0.002$& 8.162  &  0     \\
sa96-36     & 10.946 & $+0.010$& 10.836 & $+0.002$& 10.598 & $-0.007$& 10.469 & $-0.012$& 10.331 & $-0.010$ \\
sa96-180$^c$&10.875 & $-0.055 $& 10.027 & $-0.048$& 8.957  & $-0.027$& 8.410  & $-0.028$& 7.892  & $-0.013$ \\
sa96-235    & 13.097 & $+0.015$& 12.216 & $-0.002$& 11.145 & $-0.005$& 10.594 & $-0.013$& 10.069 & $+0.003$ \\
sa96-393    & 10.283 & $+0.007$& 10.261 & $-0.013$& 9.659  & $-0.007$& 9.303  & $+0.004$& 8.960  & $+0.005$ \\
sa96-406    & 9.656  & $+0.012$& 9.518  & $+0.002$& 9.306  & $-0.006$& 9.189  & $-0.005$& 9.068  & $-0.005$ \\
sa96-737    & 14.191 & $+0.019$& 13.041 & $+0.009$& 11.717 & $-0.001$& 11.002 & $-0.019$& 10.298 & $-0.010$ \\
sa97-249    & 12.487 & $-0.003$& 12.383 & $+0.003$& 11.737 & $-0.002$& 11.373 & $-0.004$& 11.021 & $-0.005$ \\
sa97-346    & 9.957  & $+0.011$& 9.863  & $-0.009$& 9.261  & $-0.001$& 8.916  & $+0.006$& 8.594  & $+0.004$ \\
sa97-351    & 10.057 & $+0.022$& 9.978  & $+0.005$& 9.786  & $-0.005$& 9.653  & $+0.004$& 9.506  & $+0.011$ \\
sa98-193    & 12.326 & $+0.036$& 11.206 & $+0.004$& 10.033 & $-0.003$& 9.442  & $-0.027$& 8.889  & $-0.012$ \\
sa98-320    & 11.451 & $+0.008$& 10.349 & $-0.026$& 9.192  & $-0.012$& 8.605  & $-0.021$& 8.079  & $-0.015$ \\
sa98-653    &\nodata &\nodata &\nodata &\nodata &\nodata &\nodata &\nodata&\nodata  & 9.523  & $-0.001$ \\
sa98-667    & 8.059  & $+0.011$& 8.394  & $+0.012$& 8.370  & $+0.008$& 8.301  & $+0.006$& 8.217  & $+0.012$ \\
sa98-978    & 11.284 & $-0.009$& 11.179 & $+0.002$& 10.572 &  0     & 10.235 & $-0.012$& 9.916  & $-0.015$ \\
sa99-6      & 3.612  & $-0.022$& 12.328 & $-0.025$& 11.070 & $-0.016$& 10.435 & $-0.033$& 9.837  & $-0.012$ \\
sa99-185    & 10.388 & $+0.012$& 9.445  & $-0.020$& 8.354  & $-0.010$& 7.799  & $-0.018$& 7.278  & $-0.005$ \\
sa99-296    & 10.890 & $+0.016$& 9.656  & $-0.015$&  8.460 & $-0.006$& 7.861  & $-0.007$& 7.322  & $+0.009$ \\
sa99-358    & 10.871 & $+0.019$& 10.383 & $-0.002$& 9.598  & $+0.007$& 9.163  & $+0.010$& 8.751  & $+0.016$ \\
sa99-408    & 10.257 &  0     & 10.223 & $-0.009$& 9.812  & $-0.005 $& 9.553  & $+0.001$& 9.306  & $+0.002$ \\
sa99-418    & 9.289  & $-0.010$& 9.429  & $+0.004$& 9.469  & $+0.005$& 9.471  & $+0.006$& 9.489  & $+0.001$ \\
sa99-438    & 8.534  & $-0.016$& 9.255  & $-0.012$& 9.396  & $+0.002$& 9.436  & $+0.021$& 9.521  & $+0.018$ \\
sa99-447    & 9.143  & $-0.018$& 9.354  & $-0.004$& 9.422  & $-0.005$& 9.460  & $-0.011$& 9.505  & $-0.014$ \\
sa100-95    & 10.111 & $+0.009$& 9.750  & $-0.021$& 8.927  & $-0.012$& 8.492  & $-0.030$& 8.058  & $-0.027$ \\
sa100-162   & 11.918 & $+0.005$& 10.444 & $-0.018$& 9.158  & $-0.008$& 8.522  & $-0.021$& 7.959  & $-0.012$ \\
sa100-241   & 10.403 & $-0.006$& 10.301 & $-0.005$& 10.151 & $-0.012$& 10.081 & $-0.020$& 9.997  & $-0.021$ \\
sa100-280   & 12.291 &  0     & 12.292 & $+0.001$& 11.803 & $-0.004$& 11.510 & $-0.006$& 11.221 & $-0.010$ \\
sa100-606   & 8.790  & $+0.028$& 8.702  & $-0.009$& 8.655  & $-0.014$& 8.635  & $-0.020$& 8.613  & $-0.020$ \\
sa101-24    & 10.127 & $+0.015$& 9.127  & $-0.022$& 8.000  & $-0.003$& 7.434  & $-0.012$& 6.914  & $-0.008$ \\
sa101-281   & 12.844 & $-0.038$& 12.405 & $-0.018$& 11.582 & $-0.007$& 11.122 & $+0.001$& 10.706 & $+0.005$ \\
sa101-282   & 10.430 & $+0.011$& 10.436 & $-0.005$& 10.005 & $-0.003$& 9.756  & $-0.014$& 9.501  & $-0.019$ \\
sa101-311   & 8.492  & $+0.009$& 8.496  & $+0.002$& 8.235  & $-0.002$& 8.082  & $-0.008$& 7.921  & $-0.010$ \\
sa101-324   & 12.031 & $+0.020$& 10.914 & $-0.011$& 9.750  & $-0.008$& 9.174  & $-0.023$& 8.643  & $-0.011$ \\
sa101-333   & 11.114 & $-0.007$& 9.355  & $-0.035$& 7.854  & $-0.019$& 7.072  & $-0.032$& 6.337  & $-0.026$ \\
sa101-363   & 10.255 & $+0.009$& 10.137 & $-0.002$& 9.882  & $-0.008$& 9.740  & $-0.012$& 9.587  & $-0.010$ \\
sa101-389   & 10.397 & $-0.009$& 10.399 & $-0.010$& 9.967  & $-0.005$& 9.707  & $-0.001$& 9.459  &  0     \\
sa102-58    & 9.470  & $-0.009$& 9.452  & $-0.012$& 9.387  & $-0.007$& 9.342  & $-0.006$& 9.336  & $-0.016$ \\
sa102-276   & 10.389 & $-0.011$& 10.409 & $-0.007$& 9.915  & $-0.005$& 9.625  & $-0.006$& 9.343  & $-0.008$ \\
sa102-381   & 8.315  & $+0.005$& 8.237  & $-0.012$& 7.930  & $-0.014$& 7.760  & $-0.017$& 7.592  & $-0.022$ \\
sa102-466   & 11.218 & $+0.005$& 10.319 & $-0.017$& 9.255  & $-0.009$& 8.701  & $-0.018$& 8.184  & $-0.007$ \\
sa102-472   & 10.579 & $+0.008$& 9.789  & $-0.021$& 8.764  & $-0.010$& 8.246  & $-0.020$& 7.755  & $-0.012$ \\
sa102-620   & 12.187 & $-0.015$& 11.196 & $-0.044$& 10.081 & $-0.012$& 9.450  & $-0.023$& 8.912  & $-0.010$ \\
sa102-625   & 9.477  &  0     & 9.454  & $-0.012$& 8.907  & $-0.017$& 8.605  & $-0.027$& 8.304  & $-0.035 $\\
sa102-1081  & 10.828 & $-0.006$& 10.581 & $-0.014$& 9.914  & $-0.011$& 9.559  & $-0.022$& 9.229  & $-0.024$ \\
sa103-302   & 10.181 & $-0.008$& 10.239 & $-0.010$& 9.868  & $-0.007$& 9.639  & $-0.006$& 9.403  & $-0.007$ \\
sa103-462   & 10.778 & $-0.014$& 10.692 & $-0.017$& 10.120 & $-0.009$& 9.794  & $-0.007$& 9.481  &  0     \\
sa103-483   & 8.870  & $+0.003$& 8.785  & $-0.005$& 8.359  & $-0.006$& 8.106  & $+0.002$& 7.869  & $+0.004$ \\
sa104-306$^b$&12.447&$+0.181$& 10.876&$+0.086$& 9.315  & $+0.055$& 8.501  & $+0.037$ &7.737  & $+0.042$ \\
sa104-337   & 12.337 & $-0.026$& 11.999 & $-0.024$& 11.217 & $-0.010$& 10.785 & $-0.012$& 10.382 & $-0.007$ \\
sa104-461   & 10.170 & $-0.019$& 10.202 & $-0.021$& 9.724  & $-0.019$& 9.427  & $-0.011$& 9.128  & $-0.003$ \\
sa104-598   & 13.610 & $+0.025$& 12.585 &  0     & 11.448 & $+0.031$& 10.832 & $-0.023$& 10.295 & $-0.031$ \\
sa105-28    & 10.250 & $+0.004$& 9.412  & $-0.028$& 8.368  & $-0.023$& 7.833  & $-0.021$& 7.332  & $-0.005$ \\
sa105-66    & 9.155  & $-0.020$& 9.131  & $-0.029$& 8.791  & $-0.031$& 8.569  & $-0.020$& 8.346  & $-0.015$ \\
sa105-205   & 11.748 & $+0.029$& 10.193 & $-0.032$& 8.811  & $-0.013$& 8.064  & $-0.010$& 7.365  & $+0.014$ \\
sa105-214   & 7.583  & $-0.003$& 7.604  & $-0.014$& 7.077  & $-0.015$& 6.759  & $-0.010$& 6.445  & $-0.007$ \\
sa105-405   & 11.703 & $+0.032$& 9.863  & $-0.033$& 8.331  & $-0.022$& 7.497  & $-0.020$& 6.683  & $+0.009$ \\
sa105-448   & 9.457  & $+0.005$& 9.437  & $-0.012$& 9.192  & $-0.016$& 9.035  & $-0.008$& 8.869  & $-0.004$ \\
sa105-663   & 11.143 & $+0.016$& 10.415 & $-0.012$& 9.432  & $-0.006$& 8.919  & $-0.015$& 8.428  & $-0.005$ \\
sa106-575   & 12.109 & $+0.023$& 10.674 & $-0.025$& 9.357  & $-0.016$& 8.691  & $-0.022$& 8.083  & $-0.007$ \\
sa106-700   & 12.678 & $+0.051$& 11.170 & $-0.023$& 9.798  & $-0.013$& 9.082  & $-0.025$& 8.419  & $-0.004$ \\
sa106-834   & 10.084 & $-0.003$& 9.798  & $-0.009$& 9.093  & $-0.005$& 8.712  & $-0.003$& 8.360  & $-0.008$ \\
sa106-1250  & 9.986  & $-0.002$& 9.180  & $-0.028$& 8.144  & $-0.021$& 7.617  & $-0.026$& 7.115  & $-0.018$ \\
sa107-35    & 10.347 & $+0.016$& 9.064  & $-0.010$& 7.786  & $-0.007$& 7.137  & $-0.021$& 6.537  & $-0.007$ \\
sa107-544   & 9.589  & $+0.005$& 9.449  & $-0.011$& 9.046  & $-0.009$& 8.812  & $-0.008$& 8.586  & $-0.006$ \\
sa107-684   & 9.136  & $-0.012$& 9.067  & $-0.015$& 8.442  & $-0.009$& 8.085  & $-0.008$& 7.733  & $-0.008$ \\
 
\enddata 
\tablenotetext{\dag} {Observed magnitudes taken from \citet{landolt83}, \citet{landolt92a}, \citet{landolt92b}, \citet{hamuy92} \& \citet{landolt99}.}
\tablenotetext{a} {\citet{landolt83} lists as possible variable.}
\tablenotetext{b} {\citet{landolt83} lists as variable.}
\tablenotetext{c} {Possible variable star.}
\end{deluxetable}

\clearpage
\begin{deluxetable} {lcc}
\tablecolumns{3}
\tablenum{5}
\tablewidth{0pc}
\tablecaption{Mean differences and standard deviations between observed 
and synthetic magnitudes \label{mean}}
\tablehead{ 
\colhead{Filter} &
\colhead{Mean Difference} &
\colhead{s. d.}}
\startdata   

$U$  &$+ 0.007$ &  0.023 \\ 
$B$  &$-0.010$ & 0.011 \\ 
$V$  &$-0.006$ & 0.006 \\ 
$R$  &$-0.008$ & 0.010 \\ 
$I$  &$-0.003$ &  0.007 \\ 

\enddata
\tablecomments{Mean values were determined utilizing
an outlier resistance algorithm.
Standard deviations were determined using a robust sigma
algorithm.}
\end{deluxetable}

\clearpage
\begin{deluxetable} {ccl}
\tablecolumns{3}
\tablenum{6}
\tablewidth{0pc}
\tablecaption{Wavelength shifts applied to Bessell passbands \label{shifts}}
\tablehead{
\colhead{Passband} &
\colhead{Shift [\AA]} &
\colhead{} }

\startdata

$U$ & 16  & blue\\
$B$ & 8.5 & red \\
$V$ & 6   & red \\
$R$ & 38  & red \\
$I$ & 5   & blue \\


\enddata
\end{deluxetable}

\clearpage
\begin{deluxetable} {cccccccccc}
\tablecolumns{5}
\tablenum{7}
\tablewidth{0pc}
\tablecaption{Normalized $UBV$ Standard Passbands \label{ubvbands}}
\label{ubv}
\tablehead{ 
\colhead{$\lambda$} &
\colhead{$U$} &
\colhead{$\lambda$} &
\colhead{$B$} &
\colhead{$\lambda$} &
\colhead{$V$}}

\startdata   
3000 & 0.000 &  3600 & 0.000 & 4700 & 0.000 \\ 
3050 & 0.034 &  3700 & 0.028 & 4800 & 0.027 \\ 
3100 & 0.113 &  3800 & 0.126 & 4900 & 0.160 \\ 
3150 & 0.236 &  3900 & 0.569 & 5000 & 0.462 \\ 
3200 & 0.376 &  4000 & 0.945 & 5100 & 0.790 \\ 
3250 & 0.525 &  4100 & 0.998 & 5200 & 0.979 \\ 
3300 & 0.662 &  4200 & 1.000 & 5300 & 1.000 \\ 
3350 & 0.770 &  4300 & 0.958 & 5400 & 0.960 \\ 
3400 & 0.855 &  4400 & 0.898 & 5500 & 0.873 \\ 
3450 & 0.913 &  4500 & 0.806 & 5600 & 0.759 \\ 
3500 & 0.958 &  4600 & 0.685 & 5700 & 0.645 \\ 
3550 & 0.983 &  4700 & 0.581 & 5800 & 0.533 \\ 
3600 & 1.000 &  4800 & 0.478 & 5900 & 0.422 \\ 
3650 & 0.997 &  4900 & 0.373 & 6000 & 0.324 \\ 
3700 & 0.987 &  5000 & 0.280 & 6100 & 0.241 \\ 
3750 & 0.947 &  5200 & 0.127 & 6200 & 0.173 \\ 
3800 & 0.851 &  5300 & 0.079 & 6300 & 0.118 \\ 
3850 & 0.713 &  5400 & 0.037 & 6400 & 0.070 \\ 
3900 & 0.526 &  5500 & 0.009 & 6500 & 0.039 \\ 
3950 & 0.334 &  5600 & 0.000 & 6600 & 0.021 \\ 
4000 & 0.175 & \nodata & \nodata & 6700 & 0.013\\  
4050 & 0.080 & \nodata & \nodata & 6800 & 0.010 \\
4100 & 0.035 & \nodata & \nodata & 6900 & 0.007 \\
4150 & 0.010 & \nodata & \nodata & 7000 & 0.000 \\
4200 & 0.000 & \nodata & \nodata & \nodata &\nodata\\ 
\enddata 
\end{deluxetable}

\clearpage
\begin{deluxetable} {cccc}
\tabletypesize{\footnotesize}
\tablecolumns{4}
\tablenum{8}
\tablewidth{0pc}
\tablecaption{Normalized $R$ \& $I$ Standard Passband \label{riband}}
\tablehead{
\colhead{$\lambda$} &
\colhead{$R$} &
\colhead{$\lambda$} &
\colhead{$I$}}
 
\startdata

5501.5 & $-$0.052   & 7001.9 & 0.000 \\         
5511.5 & $-$0.040   & 7011.9 & 0.001 \\         
5521.5 & $-$0.026   & 7021.9 & 0.002 \\          
5531.5 & $-$0.012   & 7031.9 & 0.003 \\      
5541.5 & 0.003   & 7041.9 & 0.004 \\      
5551.5 & 0.019   & 7051.9 & 0.005 \\     
5561.5 & 0.037   & 7061.9 & 0.007 \\       
5571.5 & 0.056   & 7071.9 & 0.010 \\     
5581.5 & 0.077   & 7081.9 & 0.014 \\        
5591.5 & 0.100   & 7091.9 & 0.021 \\      
5601.5 & 0.125   & 7101.9 & 0.034 \\    
5611.5 & 0.152   & 7111.9 & 0.048 \\    
5621.5 & 0.182   & 7121.9 & 0.066 \\    
5631.5 & 0.216   & 7131.9 & 0.087 \\     
5641.5 & 0.254   & 7141.9 & 0.109 \\     
5651.5 & 0.298   & 7151.9 & 0.132 \\      
5661.5 & 0.348   & 7161.9 & 0.155 \\      
5671.5 & 0.404   & 7171.9 & 0.176 \\    
5681.5 & 0.463   & 7181.9 & 0.197 \\      
5691.5 & 0.524   & 7191.9 & 0.222 \\        
5701.5 & 0.585   & 7201.9 & 0.252 \\       
5711.5 & 0.644   & 7211.9 & 0.290 \\      
5721.5 & 0.699   & 7221.9 & 0.326 \\    
5731.5 & 0.747   & 7231.9 & 0.355 \\      
5741.5 & 0.785   & 7241.9 & 0.385 \\    
5751.5 & 0.813   & 7251.9 & 0.423 \\     
5761.5 & 0.835   & 7261.9 & 0.460 \\    
5771.5 & 0.853   & 7271.9 & 0.494 \\     
5781.5 & 0.869   & 7281.9 & 0.528 \\     
5791.5 & 0.883   & 7291.9 & 0.561 \\     
5801.5 & 0.896   & 7301.9 & 0.591 \\     
5811.5 & 0.909   & 7311.9 & 0.621 \\     
5821.5 & 0.921   & 7321.9 & 0.653 \\    
5831.5 & 0.932   & 7331.9 & 0.687 \\      
5841.5 & 0.943   & 7341.9 & 0.713 \\   
5851.5 & 0.953   & 7351.9 & 0.737 \\      
5861.5 & 0.961   & 7361.9 & 0.759 \\       
5871.5 & 0.968   & 7371.9 & 0.781 \\  
5881.5 & 0.974   & 7381.9 & 0.801 \\  
5891.5 & 0.979   & 7391.9 & 0.820 \\       
5901.5 & 0.983   & 7401.9 & 0.837 \\   
5911.5 & 0.987   & 7411.9 & 0.854 \\   
5921.5 & 0.991   & 7421.9 & 0.868 \\     
5931.5 & 0.995   & 7431.9 & 0.881 \\    
5941.5 & 0.998   & 7441.9 & 0.894 \\       
5951.5 & 1.000   & 7451.9 & 0.905 \\         
5961.5 & 1.002   & 7461.9 & 0.916 \\     
5971.5 & 1.003   & 7471.9 & 0.926 \\     
5981.5 & 1.004   & 7481.9 & 0.935 \\     
5991.5 & 1.004   & 7491.9 & 0.944 \\      
6001.5 & 1.004   & 7501.9 & 0.952 \\       
6011.5 & 1.003   & 7511.9 & 0.959 \\     
6021.5 & 1.003   & 7521.9 & 0.965 \\   
6031.5 & 1.001   & 7531.9 & 0.970 \\   
6041.5 & 1.000   & 7541.9 & 0.975 \\    
6051.5 & 0.996   & 7551.9 & 0.979 \\   
6061.5 & 0.992   & 7561.9 & 0.982 \\   
6071.5 & 0.989   & 7571.9 & 0.986 \\   
6081.5 & 0.985   & 7581.9 & 0.954 \\   
6091.5 & 0.981   & 7591.9 & 0.751 \\   
6101.5 & 0.978   & 7601.9 & 0.443 \\   
6111.5 & 0.975   & 7611.9 & 0.452 \\   
6121.5 & 0.970   & 7621.9 & 0.585 \\   
6131.5 & 0.967   & 7631.9 & 0.570 \\   
6141.5 & 0.963   & 7641.9 & 0.626 \\   
6151.5 & 0.960   & 7651.9 & 0.743 \\     
6161.5 & 0.956   & 7661.9 & 0.847 \\     
6171.5 & 0.953   & 7671.9 & 0.919 \\    
6181.5 & 0.949   & 7681.9 & 0.963 \\    
6191.5 & 0.946   & 7691.9 & 0.983 \\       
6201.5 & 0.942   & 7701.9 & 0.995 \\      
6211.5 & 0.939   & 7711.9 & 0.999 \\      
6221.5 & 0.936   & 7721.9 & 0.998 \\    
6231.5 & 0.932   & 7731.9 & 0.997 \\ 
6241.5 & 0.928   & 7741.9 & 0.996 \\ 
6251.5 & 0.924   & 7751.9 & 0.996 \\ 
6261.5 & 0.920   & 7761.9 & 0.995 \\ 
6271.5 & 0.915   & 7771.9 & 0.994 \\
6281.5 & 0.911   & 7781.9 & 0.993 \\ 
6291.5 & 0.907   & 7791.9 & 0.992 \\ 
6301.5 & 0.902   & 7801.9 & 0.992 \\   
6311.5 & 0.898   & 7811.9 & 0.991 \\
6321.5 & 0.894   & 7821.9 & 0.990 \\
6331.5 & 0.889   & 7831.9 & 0.990 \\
6341.5 & 0.885   & 7841.9 & 0.989 \\
6351.5 & 0.881   & 7851.9 & 0.988 \\
6361.5 & 0.877   & 7861.9 & 0.988 \\
6371.5 & 0.873   & 7871.9 & 0.987 \\
6381.5 & 0.868   & 7881.9 & 0.986 \\
6391.5 & 0.864   & 7891.9 & 0.985 \\
6401.5 & 0.860   & 7901.9 & 0.985 \\
6411.5 & 0.856   & 7911.9 & 0.984 \\
6421.5 & 0.852   & 7921.9 & 0.983 \\
6431.5 & 0.848   & 7931.9 & 0.982 \\
6441.5 & 0.843   & 7941.9 & 0.982 \\
6451.5 & 0.839   & 7951.9 & 0.981 \\
6461.5 & 0.834   & 7961.9 & 0.980 \\
6471.5 & 0.829   & 7971.9 & 0.979 \\
6481.5 & 0.824   & 7981.9 & 0.978 \\
6491.5 & 0.819   & 7991.9 & 0.978 \\
6501.5 & 0.814   & 8001.9 & 0.977 \\
6511.5 & 0.809   & 8011.9 & 0.976 \\
6521.5 & 0.804   & 8021.9 & 0.974 \\
6531.5 & 0.798   & 8031.9 & 0.973 \\
6541.5 & 0.793   & 8041.9 & 0.972 \\
6551.5 & 0.788   & 8051.9 & 0.971 \\
6561.5 & 0.782   & 8061.9 & 0.970 \\
6571.5 & 0.776   & 8071.9 & 0.969 \\
6581.5 & 0.770   & 8081.9 & 0.967 \\
6591.5 & 0.764   & 8091.9 & 0.966 \\
6601.5 & 0.758   & 8101.9 & 0.963 \\
6611.5 & 0.752   & 8111.9 & 0.955 \\
6621.5 & 0.746   & 8121.9 & 0.949 \\
6631.5 & 0.740   & 8131.9 & 0.932 \\
6641.5 & 0.735   & 8141.9 & 0.915 \\
6651.5 & 0.731   & 8151.9 & 0.900 \\
6661.5 & 0.727   & 8161.9 & 0.884 \\
6671.5 & 0.723   & 8171.9 & 0.862 \\
6681.5 & 0.720   & 8181.9 & 0.873 \\
6691.5 & 0.717   & 8191.9 & 0.881 \\
6701.5 & 0.714   & 8201.9 & 0.883 \\
6711.5 & 0.711   & 8211.9 & 0.919 \\
6721.5 & 0.707   & 8221.9 & 0.912 \\
6731.5 & 0.703   & 8231.9 & 0.827 \\
6741.5 & 0.698   & 8241.9 & 0.863 \\
6751.5 & 0.692   & 8251.9 & 0.907 \\
6761.5 & 0.686   & 8261.9 & 0.897 \\
6771.5 & 0.679   & 8271.9 & 0.905 \\
6781.5 & 0.673   & 8281.9 & 0.872 \\
6791.5 & 0.666   & 8291.9 & 0.875 \\
6801.5 & 0.659   & 8301.9 & 0.884 \\
6811.5 & 0.653   & 8311.9 & 0.888 \\
6821.5 & 0.646   & 8321.9 & 0.883 \\
6831.5 & 0.640   & 8331.9 & 0.882 \\
6841.5 & 0.633   & 8341.9 & 0.886 \\
6851.5 & 0.618   & 8351.9 & 0.893 \\
6861.5 & 0.577   & 8361.9 & 0.887 \\
6871.5 & 0.515   & 8371.9 & 0.887 \\
6881.5 & 0.524   & 8381.9 & 0.884 \\
6891.5 & 0.538   & 8391.9 & 0.882 \\
6901.5 & 0.545   & 8401.9 & 0.879 \\
6911.5 & 0.559   & 8411.9 & 0.873 \\
6921.5 & 0.568   & 8421.9 & 0.870 \\
6931.5 & 0.570   & 8431.9 & 0.867 \\
6941.5 & 0.569   & 8441.9 & 0.861 \\
6951.5 & 0.567   & 8451.9 & 0.858 \\
6961.5 & 0.566   & 8461.9 & 0.854 \\
6971.5 & 0.564   & 8471.9 & 0.849 \\
6981.5 & 0.558   & 8481.9 & 0.844 \\
6991.5 & 0.552   & 8491.9 & 0.838 \\
7001.5 & 0.546   & 8501.9 & 0.833 \\  
7011.5 & 0.540   & 8511.9 & 0.828 \\
7021.5 & 0.534   & 8521.9 & 0.823 \\
7031.5 & 0.528   & 8531.9 & 0.818 \\
7041.5 & 0.522   & 8541.9 & 0.813 \\
7051.5 & 0.517   & 8551.9 & 0.808 \\
7061.5 & 0.512   & 8561.9 & 0.803 \\
7071.5 & 0.507   & 8571.9 & 0.798 \\
7081.5 & 0.502   & 8581.9 & 0.792 \\
7091.5 & 0.497   & 8591.9 & 0.786 \\
7101.5 & 0.492   & 8601.9 & 0.778 \\
7111.5 & 0.487   & 8611.9 & 0.770 \\   
7121.5 & 0.482   & 8621.9 & 0.761 \\   
7131.5 & 0.477   & 8631.9 & 0.751 \\   
7141.5 & 0.472   & 8641.9 & 0.741 \\  
7151.5 & 0.468   & 8651.9 & 0.731 \\   
7161.5 & 0.457   & 8661.9 & 0.719 \\  
7171.5 & 0.441   & 8671.9 & 0.708 \\   
7181.5 & 0.426   & 8681.9 & 0.695 \\   
7191.5 & 0.418   & 8691.9 & 0.682 \\   
7201.5 & 0.419   & 8701.9 & 0.667 \\   
7211.5 & 0.426   & 8711.9 & 0.652 \\      
7221.5 & 0.426   & 8721.9 & 0.635 \\      
7231.5 & 0.414   & 8731.9 & 0.618 \\      
7241.5 & 0.404   & 8741.9 & 0.601 \\    
7251.5 & 0.402   & 8751.9 & 0.583 \\    
7261.5 & 0.399   & 8761.9 & 0.564 \\     
7271.5 & 0.395   & 8771.9 & 0.545 \\      
7281.5 & 0.391   & 8781.9 & 0.527 \\   
7291.5 & 0.388   & 8791.9 & 0.507 \\   
7301.5 & 0.385   & 8801.9 & 0.488 \\    
7311.5 & 0.382   & 8811.9 & 0.468 \\     
7321.5 & 0.381   & 8821.9 & 0.448 \\     
7331.5 & 0.381   & 8831.9 & 0.427 \\      
7341.5 & 0.377   & 8841.9 & 0.405 \\     
7351.5 & 0.372   & 8851.9 & 0.384 \\   
7361.5 & 0.367   & 8861.9 & 0.361 \\   
7371.5 & 0.361   & 8871.9 & 0.340 \\      
7381.5 & 0.356   & 8881.9 & 0.320 \\   
7391.5 & 0.350   & 8891.9 & 0.300 \\   
7401.5 & 0.345   & 8901.9 & 0.281 \\    
7411.5 & 0.339   & 8911.9 & 0.263 \\   
7421.5 & 0.334   & 8921.9 & 0.246 \\   
7431.5 & 0.328   & 8931.9 & 0.227 \\      
7441.5 & 0.323   & 8941.9 & 0.210 \\   
7451.5 & 0.319   & 8951.9 & 0.192 \\     
7461.5 & 0.314   & 8961.9 & 0.171 \\      
7471.5 & 0.310   & 8971.9 & 0.152 \\       
7481.5 & 0.305   & 8981.9 & 0.139 \\     
7491.5 & 0.301   & 8991.9 & 0.120 \\   
7501.5 & 0.296   & 9001.9 & 0.110 \\ 
7511.5 & 0.292   & 9011.9 & 0.099 \\       
7521.5 & 0.288   & 9021.9 & 0.089 \\      
7531.5 & 0.284   & 9031.9 & 0.079 \\    
7541.5 & 0.279   & 9041.9 & 0.073 \\    
7551.5 & 0.275   & 9051.9 & 0.063 \\   
7561.5 & 0.271   & 9061.9 & 0.053 \\    
7571.5 & 0.267   & 9071.9 & 0.042 \\    
7581.5 & 0.254   & 9081.9 & 0.033 \\     
7591.5 & 0.199   & 9091.9 & 0.027 \\      
7601.5 & 0.116   & 9101.9 & 0.021 \\    
7611.5 & 0.112   & 9111.9 & 0.016 \\     
7621.5 & 0.144   & 9121.9 & 0.013 \\     
7631.5 & 0.139   & 9131.9 & 0.010 \\        
7641.5 & 0.148   & 9141.9 & 0.008 \\       
7651.5 & 0.173   & 9151.9 & 0.006 \\      
7661.5 & 0.194   & 9161.9 & 0.004 \\      
7671.5 & 0.208   & 9171.9 & 0.003 \\       
7681.5 & 0.214   & 9181.9 & 0.002 \\       
7691.5 & 0.215   & 9191.9 & 0.001 \\      
7701.5 & 0.214   & 9201.9 & 0.000 \\       
7711.5 & 0.211   & \nodata & \nodata \\ 
7721.5 & 0.207   & \nodata & \nodata \\
7731.5 & 0.204   & \nodata & \nodata \\ 
7741.5 & 0.200  &  \nodata & \nodata \\
7751.5 & 0.196  &  \nodata & \nodata \\
7761.5 & 0.193  &  \nodata & \nodata \\
7771.5 & 0.189  &  \nodata & \nodata \\
7781.5 & 0.185  &  \nodata & \nodata \\
7791.5 & 0.182  &  \nodata & \nodata \\
7801.5 & 0.178  &  \nodata & \nodata \\
7811.5 & 0.175  &  \nodata & \nodata \\
7821.5 & 0.171  &  \nodata & \nodata \\
7831.5 & 0.168  &  \nodata & \nodata \\
7841.5 & 0.165  &  \nodata & \nodata \\
7851.5 & 0.161  &  \nodata & \nodata \\
7861.5 & 0.158  &  \nodata & \nodata \\
7871.5 & 0.155  &  \nodata & \nodata \\
7881.5 & 0.151  &  \nodata & \nodata \\
7891.5 & 0.148  &  \nodata & \nodata \\
7901.5 & 0.145  &  \nodata & \nodata \\
7911.5 & 0.142  &  \nodata & \nodata \\
7921.5 & 0.139  &  \nodata & \nodata \\
7931.5 & 0.136  &  \nodata & \nodata \\
7941.5 & 0.133  &  \nodata & \nodata \\
7951.5 & 0.130  &  \nodata & \nodata \\
7961.5 & 0.127  &  \nodata & \nodata \\
7971.5 & 0.124  &  \nodata & \nodata \\
7981.5 & 0.121  &  \nodata & \nodata \\
7991.5 & 0.118  &  \nodata & \nodata \\
8001.5 & 0.115  &  \nodata & \nodata \\
8011.5 & 0.113  &  \nodata & \nodata \\
8021.5 & 0.110  &  \nodata & \nodata \\
8031.5 & 0.107  &  \nodata & \nodata \\
8041.5 & 0.105  &  \nodata & \nodata \\
8051.5 & 0.102  &  \nodata & \nodata \\
8061.5 & 0.100  &  \nodata & \nodata \\
8071.5 & 0.097  &  \nodata & \nodata \\
8081.5 & 0.095  &  \nodata & \nodata \\
8091.5 & 0.092  &  \nodata & \nodata \\
8101.5 & 0.090  &  \nodata & \nodata \\
8111.5 & 0.087  &  \nodata & \nodata \\
8121.5 & 0.085  &  \nodata & \nodata \\
8131.5 & 0.081  &  \nodata & \nodata \\
8141.5 & 0.078  &  \nodata & \nodata \\
8151.5 & 0.075  &  \nodata & \nodata \\
8161.5 & 0.072  &  \nodata & \nodata \\
8171.5 & 0.068  &  \nodata & \nodata \\
8181.5 & 0.067  &  \nodata & \nodata \\
8191.5 & 0.066  &  \nodata & \nodata \\
8201.5 & 0.065  &  \nodata & \nodata \\
8211.5 & 0.066  &  \nodata & \nodata \\
8221.5 & 0.064  &  \nodata & \nodata \\
8231.5 & 0.056  &  \nodata & \nodata \\
8241.5 & 0.057  &  \nodata & \nodata \\
8251.5 & 0.058  &  \nodata & \nodata \\
8261.5 & 0.056  &  \nodata & \nodata \\
8271.5 & 0.055  &  \nodata & \nodata \\
8281.5 & 0.052  &  \nodata & \nodata \\
8291.5 & 0.051  &  \nodata & \nodata \\
8301.5 & 0.050  &  \nodata & \nodata \\
8311.5 & 0.049  &  \nodata & \nodata \\
8321.5 & 0.047  &  \nodata & \nodata \\
8331.5 & 0.046  &  \nodata & \nodata \\
8341.5 & 0.045  &  \nodata & \nodata \\
8351.5 & 0.044  &  \nodata & \nodata \\
8361.5 & 0.042  &  \nodata & \nodata \\
8371.5 & 0.041  &  \nodata & \nodata \\
8381.5 & 0.040  &  \nodata & \nodata \\
8391.5 & 0.039  &  \nodata & \nodata \\
8401.5 & 0.037  &  \nodata & \nodata \\
8411.5 & 0.036  &  \nodata & \nodata \\
8421.5 & 0.035  &  \nodata & \nodata \\
8431.5 & 0.033  &  \nodata & \nodata \\
8441.5 & 0.032  &  \nodata & \nodata \\
8451.5 & 0.031  &  \nodata & \nodata \\
8461.5 & 0.030  &  \nodata & \nodata \\
8471.5 & 0.029  &  \nodata & \nodata \\
8481.5 & 0.027  &  \nodata & \nodata \\
8491.5 & 0.026  &  \nodata & \nodata \\
8501.5 & 0.025  &  \nodata & \nodata \\
8511.5 & 0.024  &  \nodata & \nodata \\
8521.5 & 0.023  &  \nodata & \nodata \\
8531.5 & 0.022  &  \nodata & \nodata \\
8541.5 & 0.021  &  \nodata & \nodata \\
8551.5 & 0.020  &  \nodata & \nodata \\
8561.5 & 0.019  &  \nodata & \nodata \\
8571.5 & 0.018  &  \nodata & \nodata \\
8581.5 & 0.017  &  \nodata & \nodata \\
8591.5 & 0.016  &  \nodata & \nodata \\
8601.5 & 0.016  &  \nodata & \nodata \\
8611.5 & 0.015  &  \nodata & \nodata \\
8621.5 & 0.014  &  \nodata & \nodata \\
8631.5 & 0.013  &  \nodata & \nodata \\
8641.5 & 0.013  &  \nodata & \nodata \\
8651.5 & 0.012  &  \nodata & \nodata \\
8661.5 & 0.012  &  \nodata & \nodata \\
8671.5 & 0.011  &  \nodata & \nodata \\
8681.5 & 0.011  &  \nodata & \nodata \\
8691.5 & 0.010  &  \nodata & \nodata \\
8701.5 & 0.010  &  \nodata & \nodata \\
8711.5 & 0.009  &  \nodata & \nodata \\
8721.5 & 0.009  &  \nodata & \nodata \\
8731.5 & 0.008  &  \nodata & \nodata \\
8741.5 & 0.008  &  \nodata & \nodata \\
8751.5 & 0.007  &  \nodata & \nodata \\
8761.5 & 0.007  &  \nodata & \nodata \\
8771.5 & 0.007  &  \nodata & \nodata \\
8781.5 & 0.006  &  \nodata & \nodata \\
8791.5 & 0.006  &  \nodata & \nodata \\
8801.5 & 0.006  &  \nodata & \nodata \\
8811.5 & 0.005  &  \nodata & \nodata \\
8821.5 & 0.005  &  \nodata & \nodata \\
8831.5 & 0.005  &  \nodata & \nodata \\
8841.5 & 0.004  &  \nodata & \nodata \\
8851.5 & 0.004  &  \nodata & \nodata \\
8861.5 & 0.004  &  \nodata & \nodata \\
8871.5 & 0.004  &  \nodata & \nodata \\
8881.5 & 0.003  &  \nodata & \nodata \\
8891.5 & 0.003  &  \nodata & \nodata \\
8901.5 & 0.003  &  \nodata & \nodata \\
8911.5 & 0.003  &  \nodata & \nodata \\
8921.5 & 0.002  &  \nodata & \nodata \\
8931.5 & 0.002  &  \nodata & \nodata \\
8941.5 & 0.002  &  \nodata & \nodata \\
8951.5 & 0.002  &  \nodata & \nodata \\
8961.5 & 0.001  &  \nodata & \nodata \\
8971.5 & 0.001  &  \nodata & \nodata \\
8981.5 & 0.000  &  \nodata & \nodata \\
8991.5 & 0.000  &  \nodata & \nodata \\
9001.5 & 0.000  &  \nodata & \nodata \\
\enddata
\end{deluxetable}

\clearpage
\begin{deluxetable} {lcc}
\tablecolumns{3}
\tablenum{9}
\tablewidth{0pc}
\tablecaption{Filter Shifts \label{nicklist}}
\tablehead{
\colhead{Passband} &
\colhead{Shift [\AA]} &
\colhead{Color Term} }
\startdata

$U$ & 12   & ($U-B$)\\
$B$ & 7.4  & ($B-V$)\\ 
$V$ & 15.2 & ($B-V$)\\
$R$ & 12.4 & ($V-R$)\\
$I$ & 40.5 & ($V-I$)\\

\enddata
\tablecomments{\small{All shifts to the red.}}
\end{deluxetable}

\clearpage
\begin{deluxetable} {lccccccccc}
\tablecolumns{10}
\tablenum{10}
\rotate
\tablewidth{0pc}
\tablecaption{Spectrophotometry of the Sun, Sirius \& Vega\label{sun}}
\tablehead{
\colhead{passband} &
\colhead{$U$} &
\colhead{$B$} &
\colhead{$V$} &
\colhead{$R$} &
\colhead{$I$} &
\colhead{$J$} &
\colhead{$H$} &
\colhead{$K$} & 
\colhead{ref.} }

\startdata

\bf{Sun}\\     
m${\rm_{obs}}$    & $-25.947$ &$-26.104$ &$-26.755$ &$-27.118$ &$-27.464$ &$-27.885$ &$-28.219$ &$-28.261$ & 1  \\
m${\rm_{syn}}$   & $-25.968$ &$-26.105$ &$-26.764$ &$-27.121$ &$-27.456$ &$-27.939$ &$-28.260$ &$-28.307$ \\
m${\rm_{obs}}$ - m${\rm_{syn}}$& $+0.021$  &$+0.001$  &$+0.009$  &$+0.003$  &$-0.008$  &$+0.054 $ &$+0.041$  &$+0.046$ \\
\tableline
\bf{Sirius} \\
m${\rm_{obs}}$&$-1.480$  &$-1.435$   &$-1.430$   &$-1.419$   &$-1.412$   &$-1.385$   &$-1.382$   &$-1.367$   & 2\\
m${\rm_{syn}}$&$-1.438$  &$-1.435$   &$-1.423$   &$-1.390$   &$-1.374$   &$-1.392$   &$-1.381$   &$-1.377$\\
m${\rm_{obs}}$ - m${\rm_{syn}}$&$-0.042$  & 0         &$-0.007$   &$-0.029$   &$-0.038$   &$+0.007$   &$-0.001$   &$+0.010$\\
\tableline
\bf{Vega} \\
m${\rm_{obs}}$&$+0.025$  &$+0.025$  &$+0.030$    &$+0.039$    &$+0.035$   &$-0.001$    &0        &$-0.001$ &  3\\
m${\rm_{syn}}$&$+0.088$  &$+0.003$  &$+0.026$    &$+0.052$    &$+0.045$   & 0          &0        &0\\
m${\rm_{obs}}$ - m${\rm_{syn}}$&$-0.063$  &$+0.022$  &$+0.004$    &$-0.013$    &$-0.010$   &$-0.001$    &0        &$-0.001$\\

\enddata

\tablerefs{
(1) Averaged values from Table A3 of \citet{bessell98} referenced from 
\citet{stebbins57}, \citet{colina96} \& \citet{cayrel96}; (2) Table A2 \citet{bessell98} 
\& references within, $UBRI$ averaged values, $JHK$ Table A1 \citet{cohen99}; 
(3) Table A2 Bessell et al. 1998, $B$ averaged value, $JHK$ Table A2 \citet{cohen99}.}

\end{deluxetable}

%

\begin{thebibliography}{}
\bibitem[Bessell(1983)]{bessell83} 
Bessell, M. S. 1983, \pasp, 95, 480

\bibitem[Bessell(1990)]{bessell90} 
Bessell, M. S. 1990, \pasp, 102, 1181

\bibitem[Bessell et al.(1998)]{bessell98} 
Bessell, M. S., Castelli, F. \& Plez,~B.\ 1998, \aap, 337, 321 (erratum)

\bibitem[Bessell(1999)]{bessell99} 
Bessell, M. S. 1999, \pasp, 111, 1426

\bibitem[Candia et al.(2003)]{candia03} 
Candia, P., et al. 2003, \pasp, 115, 277

\bibitem[Cohen et al.(1999)]{cohen99} 
Cohen, M., Walker, R. G., Carter, B., et al. 1999, \aj, 117, 1864

\bibitem[Colina et al.(1996)]{colina96} 
Colina, L., Bohlin, R. C. \& Castelli, F. 1996, \aj, 112, 307

\bibitem[Cayrel de Strobel(1996)]{cayrel96} 
Cayrel de Strobel, G. 1996, \aapr, 7, 243

\bibitem[Elias et al.(1982)]{elias82} 
Elias, J. H., Frogel, J. A., Matthews, K. \& Neugerbauer, ~G.\ 1982, \aj, 87, 1029

\bibitem[Hamuy et al.(1990)]{hamuy90} 
Hamuy, M., Suntzeff, N. B., Bravo, J. \& Phillips, M. M. 1990, \pasp, 102, 888

\bibitem[Hamuy et al.(1992)]{hamuy92} 
Hamuy, M., et al. 1992, \pasp, 104, 533

\bibitem[Hamuy et al.(1994)]{hamuy94} 
Hamuy, M., et al. 1994, \pasp, 106, 566

\bibitem[Hamuy et al.(2001)]{hamuy01} 
Hamuy, M., et al. 2001, \apj, 558, 615

\bibitem[Hayes(1970)]{hayes70} 
Hayes, D. S. 1970, \apj, 159, 165

\bibitem[Hayes(1985)]{hayes85} 
Hayes, D. S. 1985, IAU Symp.~111: Calibration of Fundamental Stellar Quantities,111, 225

\bibitem[Jha (2002)]{jha02} 
Jha, S. 2002, PhD. Thesis, Harvard University

\bibitem[Kurucz et al.(1984)]{kurucz84} 
Kurucz, R. L., et al. 1984, Solar Flux Atlas from 296 to 1300 nm (Sunspot, NM:Natl. Solar Obs.)


\bibitem[Krisciunas et al.(2003)]{krisciunas03} 
Krisciunas, K., Suntzeff, N. B., et al. 2003, \aj, 125, 166

\bibitem[Krisciunas et al.(2004)]{krisciunas04} 
Krisciunas, K., et al.  2004, \aj, 127, 1664

\bibitem[Landolt(1983)]{landolt83} 
Landolt, A. U. 1983, \aj, 88, 439

\bibitem[Landolt(1992a)]{landolt92a} 
Landolt, A. U. 1992, \aj, 104, 340

\bibitem[Landolt(1992b)]{landolt92b} 
Landolt, A. U. 1992, \aj, 104, 372

\bibitem[Landolt(1999)]{landolt99} 
Landolt, A. U. 1999, (private communication)

\bibitem[Livingston \& Wallace(1991)]{livingston91} 
Livingston, W.~\& Wallace, L.\ 1991, NSO Technical Report, Tucson: 
National Solar Observatory, National Optical Astronomy Observatory

\bibitem[Massey et al.(1988)]{massey88} 
Massey, P., et al. 1988, \apj, 328, 344

\bibitem[Menzies(1989)]{menzies89} 
Menzies, J. W. 1989, \mnras, 237, 21


\bibitem[Persson et al.(1998)]{persson98} 
Persson, S. E., Murphy, D. C., Krzeminski, W., Roth, M.~\& Rieke, M. J.\ 1998, \aj, 116, 2475

\bibitem[Pignata et al.(2004)]{pignata04}
Pignata, G., et al. 2004, \mnras, 355, 178

\bibitem[Stebbins \& Kron(1957)]{stebbins57} 
Stebbins, J., Kron, G.E.\ 1957, \apj, 126, 226

\bibitem[Stritzinger et al.(2002)]{stritzinger02} 
Stritzinger, M., Hamuy, M., Suntzeff, N.~B., et al. 2002, \aj, 124, 2100

\bibitem[Suntzeff et al.(1988)]{suntzeff88} 
Suntzeff, N. B., Hamuy, M., Martin, G., G\'omez, A. \& Gonz\'alez, R. 1988, \aj, 96, 1864

\bibitem[Suntzeff et al.(1999)]{suntzeff99} 
Suntzeff, N. B., et al. 1999, \aj, 117, 1175

\bibitem[Suntzeff(2000)]{suntzeff00} 
Suntzeff, N. B. 2000, in Cosmic Explosions, Tenth Astrophysics Conference, ed. S.S. Holt \& W.W. Zhang (AIP, New York), 65

\bibitem[Taylor(1984)]{taylor84} 
Taylor, B. J. 1984, \apjs, 54, 259


\end{thebibliography}
\end{document}